\documentclass[a4paper,12pt, epsfig]{article}
\usepackage{epsfig}
\usepackage{amssymb,latexsym}
\usepackage{amsfonts}
\usepackage{amsmath}
\usepackage[colorlinks, linkcolor=blue, citecolor=blue, urlcolor=blue]{hyperref}
\usepackage{epstopdf}
\usepackage{graphicx}
\pdfoutput=1

\newskip\humongous \humongous=0pt plus 1000pt minus 1000pt

\newif\ifdtup

\catcode`\@=11

\@addtoreset{equation}{section}
\def\theequation{\arabic{section}.\arabic{equation}}

\def\@normalsize{\@setsize\normalsize{15pt}\xiipt\@xiipt
\abovedisplayskip 14pt plus3pt minus3pt%
\belowdisplayskip \abovedisplayskip
\abovedisplayshortskip \z@ plus3pt%
\belowdisplayshortskip 7pt plus3.5pt minus0pt}

\def\small{\@setsize\small{13.6pt}\xipt\@xipt
\abovedisplayskip 13pt plus3pt minus3pt%
\belowdisplayskip \abovedisplayskip
\abovedisplayshortskip \z@ plus3pt%
\belowdisplayshortskip 7pt plus3.5pt minus0pt
\def\@listi{\parsep 4.5pt plus 2pt minus 1pt
      \itemsep \parsep
      \topsep 9pt plus 3pt minus 3pt}}

\relax

\catcode`@=12

\catcode`\@=11

\def\section{\@startsection{section}{1}{\z@}{3.5ex plus 1ex minus
    .2ex}{2.3ex plus .2ex}{\large\bf}}

\def\thesection{\arabic{section}}
\def\thesubsection{\arabic{section}.\arabic{subsection}}

\def\appendix{\setcounter{section}{0}
  \def\thesection{Appendix \Alph{section}}
  \def\thesubsection{\Alph{section}.\arabic{subsection}}
  \def\theequation{\Alph{section}.\arabic{equation}}}

\def\SymBoxes#1#2#3#4{\newdimen\un@t \un@t#3%
\raisebox{#1}{\rule{#2\un@t}{#4}\hskip-#2\un@t
\@tempdimb\un@t \advance\@tempdimb by-#4\@tempcntb#2\relax%
\@whilenum{\@tempcntb>0}\do{
\rule{#4}{\un@t}\hskip\@tempdimb \advance\@tempcntb by\m@ne}%
\hskip-#2\un@t \rule[\un@t]{#2\un@t}{#4}%
\rule[\un@t]{#4}{#4}\hskip-#4
\rule{#4}{\un@t}}\hskip-#4}                

\begin{document}


\newcommand{\dd}{\textrm{d}}

\newcommand{\beq}{\begin{equation}}
\newcommand{\eeq}{\end{equation}}
\newcommand{\bea}{\begin{eqnarray}}
\newcommand{\eea}{\end{eqnarray}}
\newcommand{\beas}{\begin{eqnarray*}}
\newcommand{\eeas}{\end{eqnarray*}}
\newcommand{\defi}{\stackrel{\rm def}{=}}
\newcommand{\non}{\nonumber}
\newcommand{\bquo}{\begin{quote}}
\newcommand{\enqu}{\end{quote}}
\newcommand{\tc}[1]{\textcolor{blue}{#1}}
\renewcommand{\(}{\begin{equation}}
\renewcommand{\)}{\end{equation}}
\def\de{\partial}
\def\Om{\ensuremath{\Omega}}
\def\Tr{ \hbox{\rm Tr}}
\def\rc{ \hbox{$r_{\rm c}$}}
\def\H{ \hbox{\rm H}}
\def\HE{ \hbox{$\rm H^{even}$}}
\def\HO{ \hbox{$\rm H^{odd}$}}
\def\HEO{ \hbox{$\rm H^{even/odd}$}}
\def\HOE{ \hbox{$\rm H^{odd/even}$}}
\def\HHO{ \hbox{$\rm H_H^{odd}$}}
\def\HHEO{ \hbox{$\rm H_H^{even/odd}$}}
\def\HHOE{ \hbox{$\rm H_H^{odd/even}$}}
\def\K{ \hbox{\rm K}}
\def\Im{ \hbox{\rm Im}}
\def\Ker{ \hbox{\rm Ker}}
\def\const{\hbox {\rm const.}}
\def\o{\over}
\def\im{\hbox{\rm Im}}
\def\re{\hbox{\rm Re}}
\def\bra{\langle}\def\ket{\rangle}
\def\Arg{\hbox {\rm Arg}}
\def\exo{\hbox {\rm exp}}
\def\diag{\hbox{\rm diag}}
\def\longvert{{\rule[-2mm]{0.1mm}{7mm}}\,}
\def\a{\alpha}
\def\b{\beta}
\def\e{\epsilon}
\def\l{\lambda}
\def\ol{{\overline{\lambda}}}
\def\ochi{{\overline{\chi}}}
\def\th{\theta}
\def\s{\sigma}
\def\oth{\overline{\theta}}
\def\ad{{\dot{\alpha}}}
\def\bd{{\dot{\beta}}}
\def\oD{\overline{D}}
\def\opsi{\overline{\psi}}
\def\dag{{}^{\dagger}}
\def\tq{{\widetilde q}}
\def\L{{\mathcal{L}}}
\def\p{{}^{\prime}}
\def\W{W}
\def\N{{\cal N}}
\def\hsp{,\hspace{.7cm}}
\def\hspp{,\hspace{.5cm}}
\def\bo{\ensuremath{\hat{b}_1}}
\def\bfo{\ensuremath{\hat{b}_4}}
\def\co{\ensuremath{\hat{c}_1}}
\def\cfo{\ensuremath{\hat{c}_4}}
\def\th#1#2{\ensuremath{\theta_{#1#2}}}
\def\c#1#2{\hbox{\rm cos}(\th#1#2)}
\def\s#1#2{\hbox{\rm sin}(\th#1#2)}
\def\cp#1#2#3{\hbox{\rm cos}^#1(\th#2#3)}
\def\sp#1#2#3{\hbox{\rm sin}^#1(\th#2#3)}
\def\ctp#1#2#3{\hbox{\rm cot}^#1(\th#2#3)}
\def\cpp#1#2#3#4{\hbox{\rm cos}^#1(#2\th#3#4)}
\def\spp#1#2#3#4{\hbox{\rm sin}^#1(#2\th#3#4)}
\def\t#1#2{\hbox{\rm tan}(\th#1#2)}
\def\tp#1#2#3{\hbox{\rm tan}^#1(\th#2#3)}
\def\m#1#2{\ensuremath{\Delta M_{#1#2}^2}}
\def\mn#1#2{\ensuremath{|\Delta M_{#1#2}^2}|}
\def\u#1#2{\ensuremath{{}^{2#1#2}\mathrm{U}}}
\def\pu#1#2{\ensuremath{{}^{2#1#2}\mathrm{Pu}}}
\def\meff{\ensuremath{\Delta M^2_{\rm{eff}}}}
\def\an{\ensuremath{\alpha_n}}
\newcommand{\Z}{\ensuremath{\mathbb Z}}
\newcommand{\R}{\ensuremath{\mathbb R}}
\newcommand{\rp}{\ensuremath{\mathbb {RP}}}
\newcommand{\vac}{\ensuremath{|0\rangle}}
\newcommand{\vact}{\ensuremath{|00\rangle}                    }
\newcommand{\oc}{\ensuremath{\overline{c}}}
\renewcommand{\cos}{\textrm{cos}}
\renewcommand{\sec}{\textrm{sec}}
\renewcommand{\sin}{\textrm{sin}}
\renewcommand{\cot}{\textrm{cot}}
\renewcommand{\tan}{\textrm{tan}}
\renewcommand{\ln}{\textrm{ln}}

\renewcommand{\re}{\ensuremath{\mathcal{E}}}

\newcommand{\Vol}{\textrm{Vol}}

\newcommand{\half}{\frac{1}{2}}

\def\changed#1{{\bf #1}}

\begin{titlepage}

\def\thefootnote{\fnsymbol{footnote}}

\begin{center}
{\large {\bf
Vetoing Cosmogenic Muons in A Large Liquid Scintillator
  } }

\bigskip

\bigskip

{\large \noindent   Marco Grassi$^{1,2}$\footnote{\texttt{mgrassi@ihep.ac.cn}}, Jarah Evslin$^{3}$\footnote{\texttt{jarah@impcas.ac.cn}}, Emilio
Ciuffoli$^{3}$\footnote{emilio@impcas.ac.cn} and Xinmin Zhang$^{4,5}$\footnote{\texttt{xmzhang@ihep.ac.cn}} }
\end{center}

\renewcommand{\thefootnote}{\arabic{footnote}}

\vskip.7cm

\begin{center}
\vspace{0em} {\em 
1) Institute of High Energy Physics (IHEP), CAS, Beijing 100049, China\\
2) INFN Sezione di Milano, via Celoria 16, 20133 Milan, Italy\\
3)  Institute of Modern Physics, CAS, NanChangLu 509, Lanzhou 730000, China\\
4) Theoretical Physics Division, IHEP, CAS, 
Beijing 100049, China\\
5) Theoretical Physics Center for Science Facilities, IHEP, CAS,
Beijing 100049, China\\

 {}}


\vskip .4cm

\vskip .4cm

\end{center}

\vspace{1.3cm}

\noindent
\begin{center} {\bf Abstract} \end{center}

\noindent
At upcoming medium baseline reactor neutrino experiments the spallation $^9$Li background will be somewhat larger than the inverse $\beta$ decay reactor neutrino signal.   We use new FLUKA simulations of spallation backgrounds to optimize a class of veto strategies and find that surprisingly the optimal veto for the mass hierarchy determination has a rejection efficiency below 90\%. The unrejected background has only a modest effect on the physics goals.  For example $\Delta\chi^2$ for the hierarchy determination falls by 1.4 to 3 points depending on the muon tracking ability.  The optimal veto strategy is essentially insensitive to the tracking ability, consisting of 2 meter radius, 1.1 second cylindrical vetoes of well tracked muons with showering energies above 3 to 4 GeV and 0.7 second full detector vetoes for poorly tracked muons above 15 to 18 GeV.  On the other hand, as the uncertainty in $\theta_{12}$ will be dominated by the uncertainty in the reactor neutrino spectrum and not statistical fluctuations, the optimal rejection efficiency for the measurement of $\theta_{12}$ is 93\% in the case of perfect tracking.


\vfill

\begin{flushleft}
{\today}
\end{flushleft}
\end{titlepage}

\hfill{}


\setcounter{footnote}{0}

\section{Introduction} \label{intro}
\noindent
The next generation of medium baseline reactor neutrino experiments \cite{juno,reno50} will deliver by far the most precise measurement yet of $\theta_{12}$ \cite{idea12} and also will be sensitive to the neutrino mass hierarchy \cite{petcovidea} .  Cosmogenic muons interacting with ${}^{12}$C in their organic liquid scintillator detectors will create spallation products ${}^9$Li and ${}^8$He whose decays yield the same double coincidence used to detect reactor neutrinos \cite{cr}.  The background will be larger than the signal \cite{noimuoni}.  The effectiveness of a veto strategy depends critically on the ability to track these muons, so that only events near the track need be vetoed.  In this note we present the optimal strategy as a function of the tracking efficiency.  Our main result is that the optimal strategy leads to the acceptance of a much larger fraction of the background than had been found in previous studies but that this large background has a relatively minor effect on the main science goals of these experiments.

On a smaller scale, this issue has already been confronted by KamLAND \cite{kamlandback}.  As the detector is deeper and smaller, the muon rate is only 0.2 Hz, appreciably slower than the 257 ms decay time of ${}^9$Li.  Of these muons, the most dangerous are the showering muons.  These are defined to be muons which deposit at least 3 GeV in the detector in addition to the energy deposited by ionization. While they occur at a rate of only 0.03 Hz, they are responsible for the majority of the ${}^9$Li background~\cite{dwyer}.  As a result, KamLAND chose a full detector veto of 2 seconds after all showering muons and also after muons whose track could not be reconstructed.  For nonshowering muons whose track could be reconstructed, a 3 meter radius cylinder about the track was vetoed for 2 seconds.  The combination of these vetoes left a small number of background events, estimated for example as 14 \cite{kamlandparam} of the original $10^3$ \cite{kamlandback}.  Overall, this strategy led to a rejection efficiency of about 99\% while removing only about 15\% of the signal.

Can this same strategy be repeated with upcoming experiments like JUNO and RENO 50?  JUNO will have a muon rate of about 5 Hz and a showering muon rate of about 1 Hz, and so a 2 second veto after each showering muon would eliminate essentially all of the events \cite{noimuoni}.  However we will show in this paper that a much less efficient veto can give an acceptable dead time and result in a relatively modest effect on the main science goals of the experiment, the determination of the neutrino mass hierarchy and the most accurate ever measurement of $\theta_{12}$.  The reason for this is that the shape and normalization of the ${}^9$Li decay spectrum can be determined fairly well, and the ${}^8$He decays yield a small background.  Therefore the background can be subtracted, leaving only a small contribution to the statistical fluctuations.  In this note we will determine the optimal veto strategy and its impact on the sensitivity to the hierarchy and the measurement of $\theta_{12}$.

This procedure is somewhat complicated by muon bundles.  At both KamLAND and JUNO, nearly half of the muons that hit the detector are contained in bundles of multiple muons created in the same cosmic ray collision.  However, as the KamLAND detector is much smaller than the characteristic muon separation, in less than 5\% of events do multiple muons from the same bundle strike the detector.  On the other hand, in the case of JUNO, about 10\% of muon events will consist of multiple muons striking the detector.  The tracking of these bundle events is extremely challenging, and so the ability to veto only the muon track, and not the full detector, will depend on whether JUNO is capable of tracking multiple muons. In this study we will examine the dependence of the sensitivity to the hierarchy and the precision with which $\theta_{12}$ can be measured upon the tracking ability of the detector.  

\section{The ${}^9$Li Background}
\noindent
51\% of ${}^{9}$Li and 16\% of ${}^8$He decays yield a neutron together with an electron, mimicking the double coincidence signal of inverse $\beta$ decay (IBD).  At JUNO these double coincidence background rates will be $8.5\times 10^{-4}$ Hz and $3.0\times 10^{-5}$ Hz respectively.  The two decay spectra are so similar that so far KamLAND has not been able to conclusively observe ${}^8$He beyond the 1 $\sigma$ level.  Therefore, to a reasonable approximation the ${}^8$He background may be ignored.  

\subsection{The Distribution of ${}^9$Li} \label{lisez}

At KamLAND a full detector or cylindrical veto was applied to all cosmogenic muons.   However those muons whose showering energy, defined to be the energy that they deposit in the scintillator minus 1.43 MeV per cm of their track\footnote{1.43 MeV/cm is the ionization energy deposited in electrons with less than 100 keV.  Those with more energy are considered to be $\delta$ rays and little would be gained by subtracting them as well from the showering energy \cite{noimuoni}.}, is sufficiently small are exceedingly unlikely to generate any ${}^9$Li.   In fact, we will see below that in the case of JUNO, there are so many muons that the optimal strategy is not to apply any veto to muons whose showering energy is less than a certain value, which is at least about 3 GeV in each case.  The average muon track length is about 24 meters.  To determine the volume-weighted dead time, we will need the average track length of only those muons whose showering energies exceed a certain minimum $E_t$.  We have found that this average track length is 28 meters whenever $E_t>1.5$ GeV, as it for the veto strategies that we will find to be optimal.  Therefore, while we will use the average track length $d(E_t)$ as determined by our FLUKA simulations, in practice for the optimal veto strategies this will always be equal to 28 meters.


\begin{figure} 
\begin{center}
\includegraphics[width=4.2in]{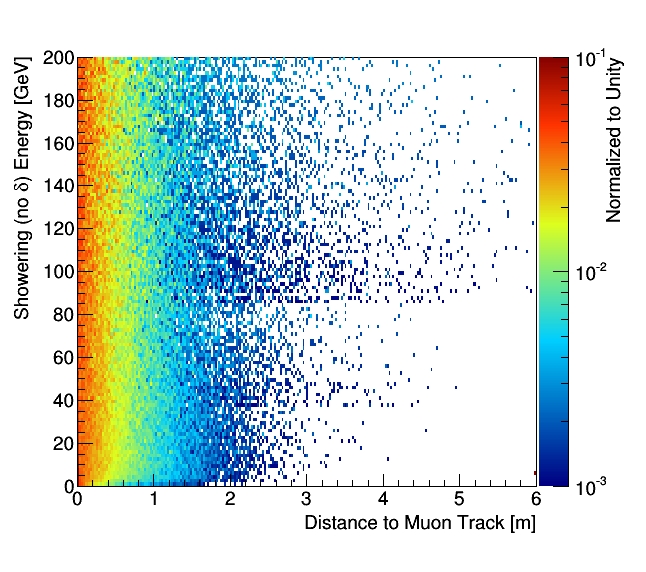}
\caption{The radial distribution (horizontal) of ${}^9$Li about a muon track, normalized to unity for each muon showering energy (vertical)}
\label{fattorizzafig}
\end{center}
\end{figure}

\begin{figure} 
\begin{center}
\includegraphics[width=3.5in]{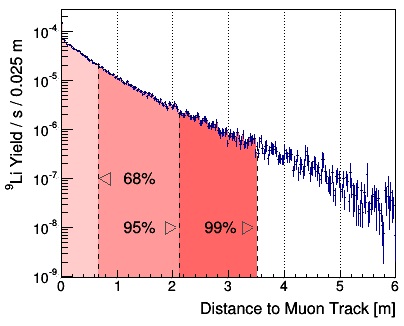}
\caption{The radial distribution of ${}^9$Li about a muon track}
\label{radfig}
\end{center}
\end{figure}

The distribution of ${}^9$Li events about this track then depends on the muon's showering energy and also the distance from the muon track.  As can be seen in Fig. \ref{fattorizzafig}, this distribution to a very good approximation factorizes, in the sense that the radial distribution of ${}^9$Li is essentially independent of the showering energy.  This greatly simplifies our analysis, as it means that the ${}^9$Li yield is entirely characterized by just two functions, the radial, shown in Fig. \ref{radfig} and showering energy distribution which was already provided in Ref.~\cite{noimuoni}.

\begin{figure} 
\begin{center}
\includegraphics[width=3.5in]{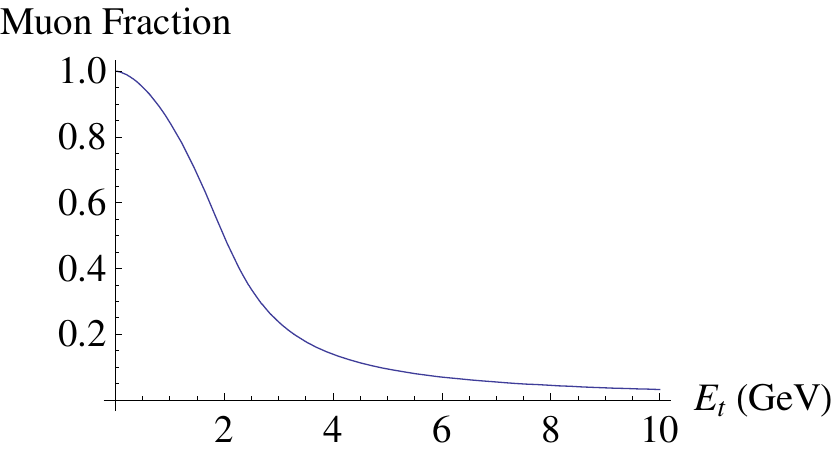}
\caption{The fraction of muons with a showering energy above the threshold $E_t$ GeV}
\label{muonfracfig}
\end{center}
\end{figure}

\begin{figure} 
\begin{center}
\includegraphics[width=3in]{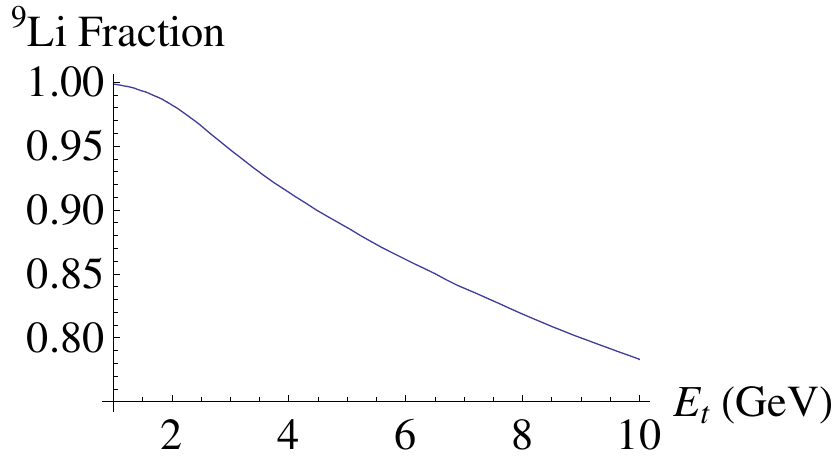}
\includegraphics[width=3in]{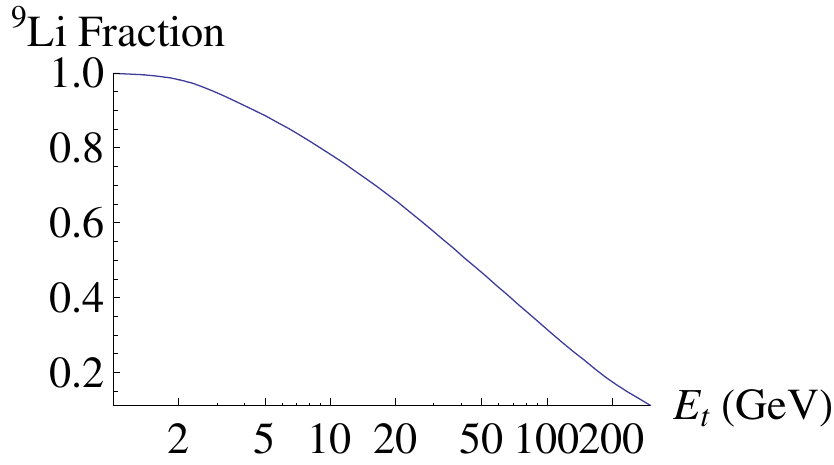}
\caption{The fraction of ${}^9$Li generated by muons with a showering energy above the threshold $E_t$ GeV}
\label{lifracfig}
\end{center}
\end{figure}

In the next sections we will attempt to determine the optimal veto strategy.  To do this, we will need the fraction of muons with a showering energy above a threshold $E_t$ and the fraction of ${}^9$Li produced by those muons.  These are displayed respectively in Figs. \ref{muonfracfig} and \ref{lifracfig}. 

\section{Live Time vs Rejection Efficiency} \label{taglisez}

\subsection{The veto strategy}

In practice the optimal veto strategy may be quite complicated, and any strategy should be refined once data taking begins.  Our goal in this paper is to understand the effect of the spallation isotope backgrounds on the science goals of medium baseline reactor experiments and the implications of these backgrounds for the muon tracking requirements.  As a result, it is necessary to produce a reasonably effective but simple veto strategy, which can be used in the analysis.

The strategies that we will consider consist of two cuts
\newcounter{cuts}
\begin{list}{\arabic{cuts})} {\usecounter{cuts} \setlength{\leftmargin}{0cm}\setlength{\itemsep}{.4cm}}
\item {\bf When the energy $E>E_f$ and it is known that the muon (bundle) was not tracked successfully, veto the full detector for a time $t_f$.}
\item {\bf When the $E>E_c$ and it is believed that the muon was tracked successfully, veto a cylinder of radius $r_c$ about each muon track for a time $t_c$.}
\end{list}
We will optimize $E_f,\ E_c,\ r_c,\ t_f$ and $t_c$ separately for the neutrino mass hierarchy determination and for the measurement of $\theta_{12}$.  We will first assume that each muon is tracked perfectly.  This assumption will be relaxed in Sec.~\ref{trackcatt}.

One might think that the simultaneous optimization of these 5 variables would require a slow, global search through a 5-dimensional space, rerunning our simulations at each step.  In fact the simulations only depend on two parameters, the background rejection efficiency and the live time.  Therefore we will use the following strategy.  First, in this section, using the results of Sec.~\ref{lisez} we will find the strategy which maximizes the live time for {\it each} possible background rejection efficiency.  This maximum is independent of the science goal considered, and so this section will apply to both the hierarchy search and to $\theta_{12}$, and can in the future be applied to other potential science goals such as the study of geoneutrinos should the reactors all be turned off in the future.  

Next, in Secs.~\ref{optsez} and \ref{trackcatt} simulations of the detector JUNO with respectively perfect and imperfect tracking are considered with each possible background rejection efficiency and live time.  As the optimal live time for each detector efficiency will already be determined, for each science goal we will only need to optimize a single variable, the background rejection efficiency.  This will yield the optimal cuts as well as the effect of the background upon the hierarchy determination and the measurement of~$\theta_{12}$.


Let $m(E)$ and $l(E)$ respectively be the fraction of single muons and the fraction of ${}^9$Li isotopes created by single muons with showering energy greater than $E$.  These two functions, in the case of the detector JUNO are plotted in Figs.~\ref{muonfracfig} and~\ref{lifracfig}.  Here we have not considered muon bundle events.   While only 10\% of muon events at JUNO will be multimuon events, these will account for about a third of all muons and so about a third of spallation isotope production.

In the case of muon bundles, the spallation isotope production rate is just the sum of the rates for the individual muons.  The energy distribution for muon bundles is similar to that of single muons and so an extension of our analysis to bundle events would be straightforward.  One need only modify the showering energy distribution $m(E)$ and spallation isotope distribution $l(E)$ functions to be the weighted sum of the distributions for all $k$-muon bundles, where the total deposited energy is the total of that of the $k$ muons.  This requires no new simulations, as the $k$-bundle distribution is essentially just the sum of $k$ single muon distributions.  One only needs to know the probability of multimuon events, which in the case $k=2$ is calculated in Ref.~\cite{noimuoni}.  

The inclusion of bundles  would reduce the rejection efficiency as the scintillator detectors cannot determine how much energy was deposited by each muon, and so one would need to rely upon a total energy threshold.  As a result, the rejection efficiencies obtained in this study are optimistic estimates.  However, it will be clear below that even a 10\% reduction in the rejection efficiency would not have an enormous impact on the science goals.


\subsection{Full detector veto}

A full detector veto of time $t_f$ after the passage of each muon with a showering energy of at least $E_f$ will yield a rejection efficiency of
\beq
\re_f(E_f,t_f)=l(E_f)(1-e^{-t_f/t_d}) \label{pienor}
\eeq
where $t_d\sim 255$ ms is the decay time for ${}^9$Li.  Only $\beta$ decays of ${}^9$Li to excited states of ${}^9$Be eventually yield free neutrons and so a false double coincidence, however the decay of ${}^9$Be into two $\alpha$ particles and a neutron is much faster than the $\beta$ decay of ${}^9$Li and the $\beta$ decay time of ${}^9$Li is independent of whether the ${}^9$Be is in an excited state, and so $t_f$ for the decays producing neutrons is equal to the overall $t_f$ for ${}^9$Li.

\begin{figure} 
\begin{center}
\includegraphics[width=3.5in]{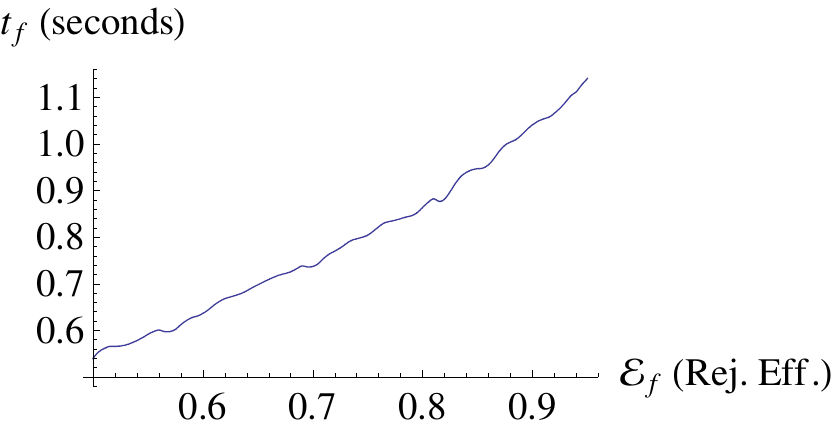}
\caption{The optimal full detector veto time $t_f(\re_f)$ yielding an ${}^{9}$Li rejection efficiency $\re_f$}
\label{tfpfig}
\end{center}
\end{figure}

For a given rejection efficiency $\re_f$, one can invert (\ref{pienor}) to obtain $E_f$ as a function of the veto time $t_f$
\beq
E_f(\re_f,t_f)=l^{-1}\left(\frac{\re_f}{1-e^{-t_f/t_d}}\right) . \label{pienoe}
\eeq
The resulting fractional live time $t_l$ is determined from the product of the veto rate $m(E_f)$ and the veto time $t_f$ 
\beq 
t_l(\re_f,t_f)={\rm{exp}}\left(-t_f\ m(E_f(\re_f,t_f))\right). \label{pienot}
\eeq
The veto may then be optimized, for each $\re_f$, by simply setting
\beq
\frac{\partial t_l(\re_f,t_f)}{\partial t_f}=0. \label{pienopar}
\eeq
The solution to this equation is the optimal $t_f$ for each rejection efficiency $\re_f$, which we will call $t_f(\re_f)$.  The optimal veto time $t_f(\re_f)$ for each efficiency $\re_f$ is shown in Fig.~\ref{tfpfig}.  Note that the optimal veto time is about 1 second, much less than the 2 seconds at KamLAND.  

\begin{figure} 
\begin{center}
\includegraphics[width=3in]{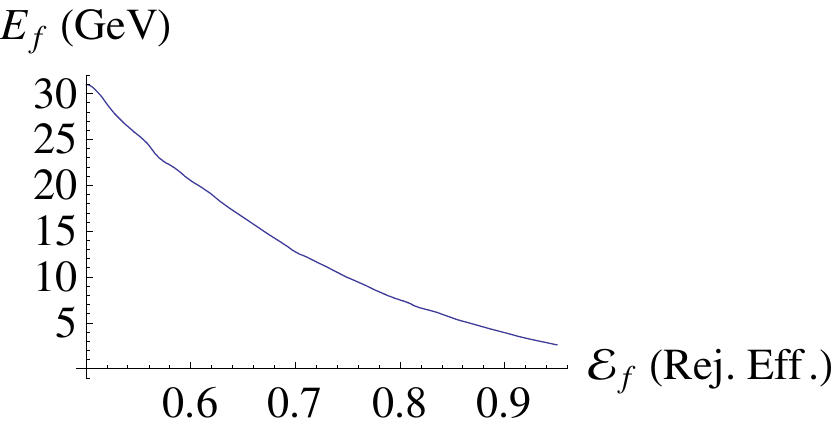}
\includegraphics[width=3in]{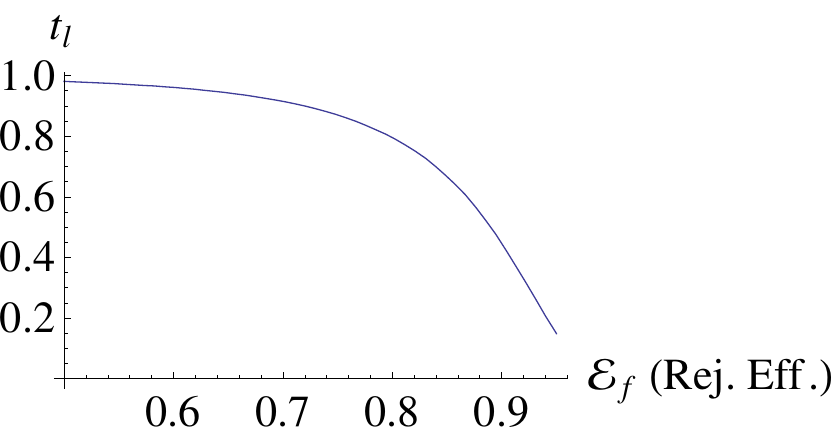}
\caption{The optimal full detector showering energy threshold $E_f(\re_f)$ and fractional live time $t_l(\re_f)$ yielding an ${}^{9}$Li rejection efficiency $\re_f$}
\label{efpfig}
\end{center}
\end{figure}

Once $t_f(\re_f)$ is found numerically, $E_f(\re_f)$ is found using Eq.~(\ref{pienoe}) and the fractional live time $t_l(\re_f)$ using~(\ref{pienot}).   These are shown in Fig.~\ref{efpfig}.  Note that even a modest rejection efficiency of $90\%$ requires a showering energy threshold of about $4$ GeV and leads to a live time of less than one half of the total runtime.   In a 20 kton detector, full detector vetoes are quite costly.  As a result, we will assume that they are only used when the muon was not tracked successfully, a case which will be considered in Sec.~\ref{trackcatt}.  In Sec.~\ref{optsez} we will assume that all muons are well tracked and so only cylindrical vetoes will be considered.

\subsection{Cylindrical veto}

The efficiency of a cylindrical veto depends not only on the functions $m(E)$ and $l(E)$ characterizing the muon distribution and ${}^9$Li yields as functions of energy $E$, but also upon the ${}^9$Li radial distribution about the muon track $f(R)$.  Here $f(R)$ is the ${}^9$Li rate per unit radius at a radial distance $R$ from the track, as illustrated in Fig.~\ref{radfig}.  A radius $r_c$ cylindrical veto of time $t_c$ after the passage of each muon with a showering energy of at least $E_c$ will yield a rejection efficiency of
\beq
\re_c(E_c,t_c,r_c)=l(E_c)(1-e^{-t_c/t_d})\frac{\int_0^{r_c} f(R)dR}{\int_0^\infty f(R)dR}. \label{cilr}
\eeq
For simplicity we will define the integral
\beq
F(r_c)=\frac{\int_0^{r_c} f(R)dR}{\int_0^\infty f(R)dR}
\eeq
which is equal to the normalized ${}^9$Li event rate at radii below $r_c$.

Note that the expression~(\ref{cilr}) for the rejection efficiency $\re_c$ factorizes into a part dependent upon $E_c$ and a part $x$ dependent upon $t_c$ and $r_c$
\beq
\re_c(E_c,t_c,r_c)=l(E_c)x(t_c,r_c)\hsp
x(t_c,r_c)=\left(1-e^{-t_c/t_d}\right)F(r_c). \label{fatte}
\eeq
Therefore we may separate the optimization problem into steps.  First we may fix $x$.  This gives a relation between $t_c$ and $r_c$
\beq
t_c(x,r_c)=-t_d\ {\rm{ln}}\left(1-\frac{x}{F(r_c)}\right) . \label{tceq}
\eeq
Now instead of the three independent variables $(E_c,t_c,r_c)$ there are three independent variables $(E_c,r_c,x)$.  However the rejection efficiency $\re_c$ is entirely determined by $E_c$ and $x$ via Eq.~(\ref{fatte}) and so $r_c$ can be varied, fixing $E_c$ and $x$, so as to minimize the total dead time.  Here the volume-weighted dead time that is relevant is the veto time multiplied by the veto volume corresponding to a live time of
\beq
t_l(E_c,r_c,x)={\rm{exp}}\left(-\frac{\pi d(E_c) r_c^2}{V} t_c(x,r_c) m(E_c)\right) \label{ciltl}
\eeq
where $V$ is the detector volume and, following the definition in Sec.~\ref{lisez}, $d(E_c)$ is the average tracklength of a muon with showering energy $E_c$.  The detector efficiency will be proportional to~$t_l$.

\begin{figure} 
\begin{center}
\includegraphics[width=3in]{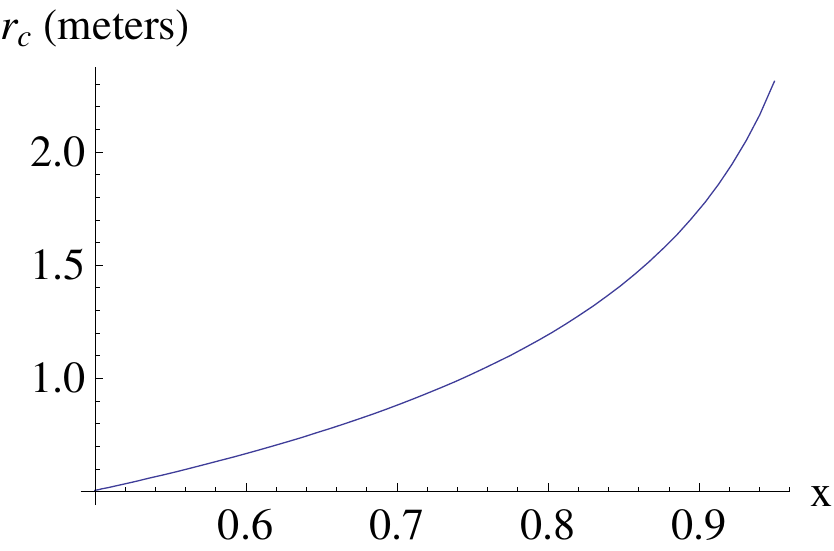}
\includegraphics[width=3in]{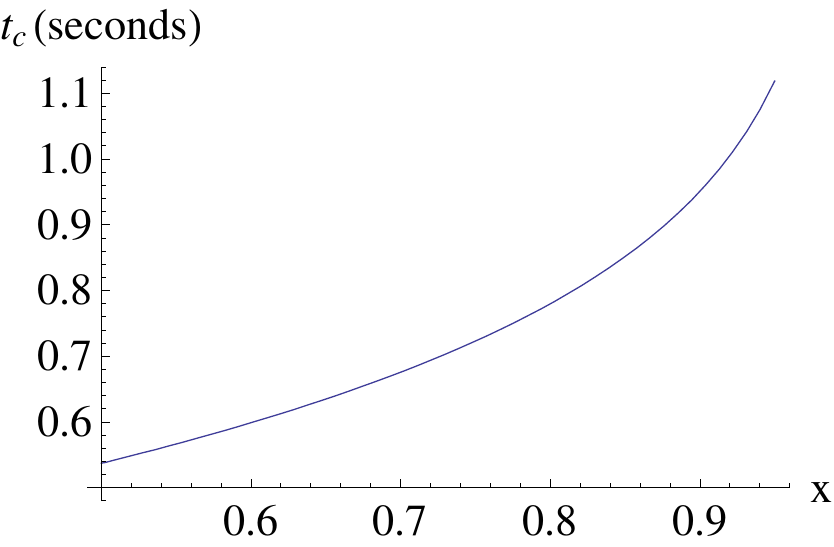}
\caption{The optimal cylindrical veto radius $r_c$ and time $t_c$ for a fixed value of $x$, defined in Eq.~(\ref{fatte})}
\label{cilrtfig}
\end{center}
\end{figure}

We want to choose $r_c$ so as to maximize $t_l$.  However $r_c$ only enters in the combination $r_c^2 t_c(x,r_c)$ and so we only need minimize $r_c^2 t_c(x,r_c)$ for each value of x
\beq
\frac{\partial r_c^2 t_c(x,r_c)}{\partial r_c}=0.
\eeq
Using Eq.~(\ref{tceq}) this minimization is easily performed, yielding the optimal value of $r_c(x)$ and therefore $t_c(x)$ for each value of $x$.  The optimal values are shown in Fig.~\ref{cilrtfig}.  Similarly to the full detector vetoes, the optimal veto time is about 1 second, only half as long as the KamLAND vetoes.  Similarly the optimal veto radius is always less than the 3 meter KamLAND veto.  Due to KamLAND's greater depth and smaller size, the muon rate is much lower which allows KamLAND to use longer vetoes with a small effect on the measurement of the reactor mixing parameters.  While an analysis such as ours applied to KamLAND would nonetheless find that the optimal veto would be longer than at JUNO but shorter than 2 seconds, nonetheless as a result of the low muon rate KamLAND does not need a precise optimization of the veto time.  On the other hand, at JUNO the conflicting constraints imposed by the large background, which requires a long veto, and the statistical fluctuations in the hierarchy-sensitive region of the spectrum around 3 MeV, which require a short veto, leave a small window of acceptable veto strategies for the hierarchy determination at JUNO.  

\begin{figure} 
\begin{center}
\includegraphics[width=3.5in]{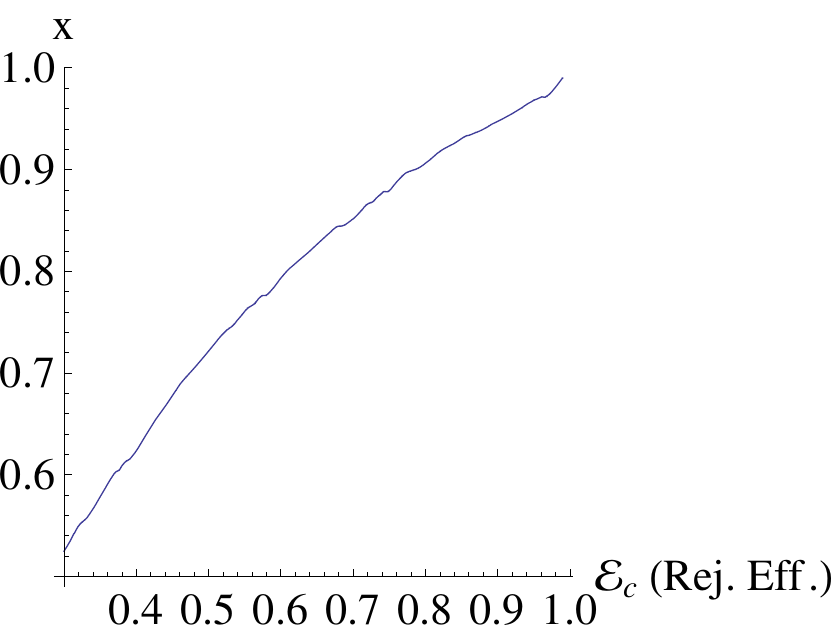}
\caption{The optimal value of $x$, defined in Eq.~(\ref{fatte}), as a function of the efficiency $\re_c$ of a cylindrical veto}
\label{cilxfig}
\end{center}
\end{figure}

We have reduced the number of independent parameters from 3 to 2, only $E_c$ and $x$ need to be optimized.  The problem is now identical to that of the full detector veto treated above but now $x$ plays the role that was played by $t_f$, or strictly speaking by $1-e^{-t_f/t_l}$.  Fixing the rejection efficiency $\re_c$, Eq.~(\ref{fatte}) gives the showering energy threshold $E_c$
\beq
E_c(\re_c,x)=l^{-1}\left(\frac{\re_c}{x}\right). \label{cilec}
\eeq
Substituting this value of $E_c(\re_c,x)$ into (\ref{ciltl}) one obtains $t_l(\re_c,x)$.  Thus, as in the case of a full detector veto (\ref{pienopar}), for each rejection efficiency $\re_c$ one determines the optimal $x(\re_c)$ by demanding that it extremizes $t_l(\re_c,x)$ with $\re_c$ held fixed.  This optimum is shown in Fig.~\ref{cilxfig}.

\begin{figure} 
\begin{center}
\includegraphics[width=3.5in]{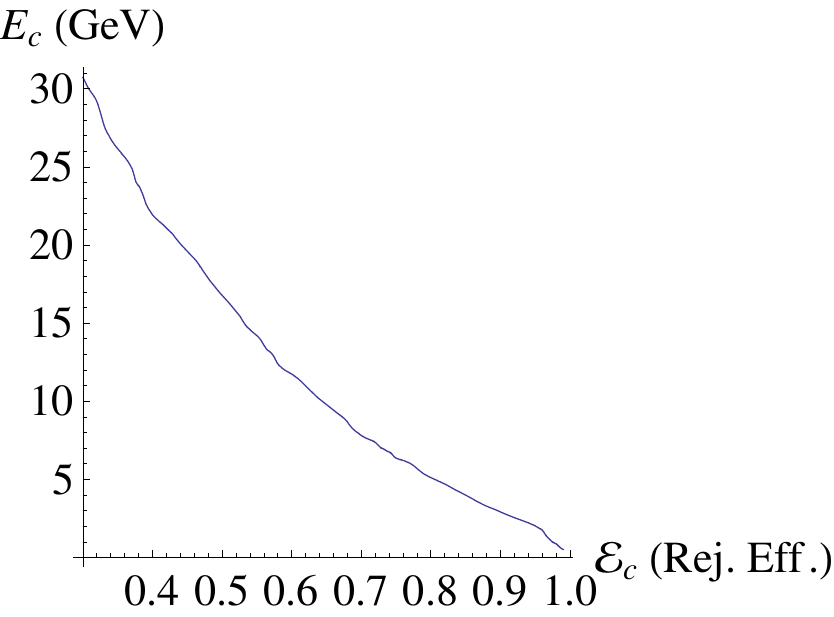}
\caption{The optimal showering energy threshold $E_c$, as a function of the efficiency $\re_c$ of a cylindrical veto}
\label{cilefig}
\end{center}
\end{figure}

Once $x(\re_c)$ has been determined, Eq.~(\ref{cilec}) yields the threshold showering energy $E_c$ for the cylindrical veto, shown in Fig.~\ref{cilefig}.  For an efficiency $\re_c$ near 99\% this threshold should be quite low, near 1 GeV.  However a more modest efficiency of 90\% requires a threshold of 3 GeV, equal to the threshold used by KamLAND for its {\it{full}} detector veto.  

\begin{figure} 
\begin{center}
\includegraphics[width=3in]{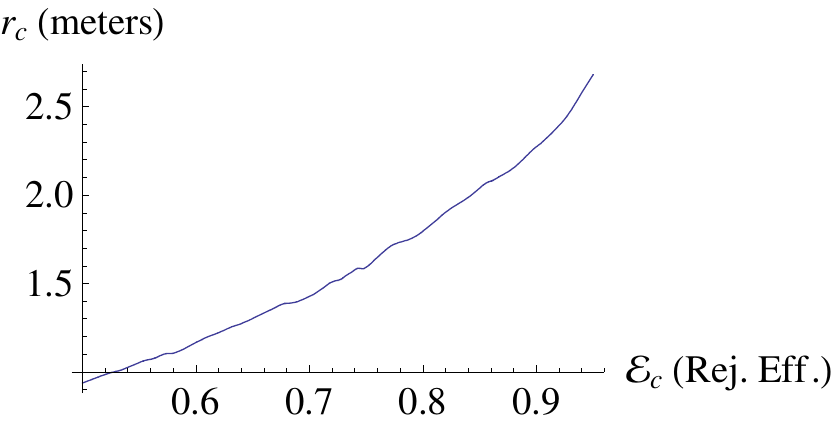}
\includegraphics[width=3in]{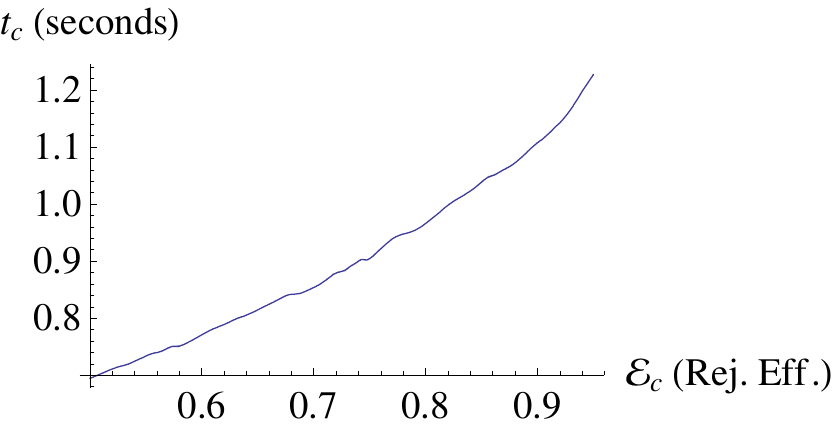}
\caption{The optimal cylindrical veto radius $r_c$ and time $t_c$ for a fixed rejection effiency $\re_c$}
\label{cilrt2fig}
\end{center}
\end{figure}

\begin{figure} 
\begin{center}
\includegraphics[width=3.5in]{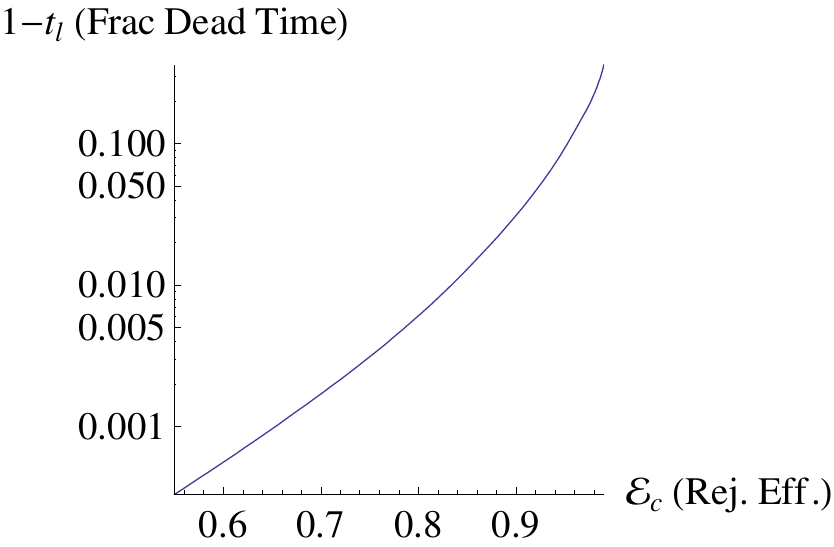}
\caption{The fractional volume-weighted dead time $1-t_l$, as a function of the efficiency $\re_c$ of a cylindrical veto}
\label{ciltlfig}
\end{center}
\end{figure}

Combining Fig.~\ref{cilxfig} for $x$ as a function of $\re_c$ with Fig.~\ref{cilrtfig} which provides $r_c$ and $t_c$ as a function of $x$ one finds the cylindrical veto parameters $r_c$ and $t_c$ as a function of the efficiency $\re_c$.  These are shown in Fig.~\ref{cilrt2fig}.  Finally, as the fractional live time is now quite close to unity, we plot the fractional dead time $1-t_l$ in Fig.~\ref{ciltlfig}.  This is one of the main results of our study.  Note that a rejection efficiency of 90\% can be obtained while losing only 3\% of the IBD signal, while a 95\% efficiency costs 10\% of the signal.  However the commonly quoted goal of 99\% background rejection would require a loss of 35\% of the IBD events, and this under the assumption of perfect tracking which is quite unlikely for 99\% of muons.

\section{Optimal Veto for Each Science Goal} \label{optsez}

In Sec.~\ref{taglisez} for each kind of veto, full detector or cylindrical, we obtained the maximum fractional live time $t_l$ possible for each ${}^9$Li rejection efficiency $\re_f$ or $\re_c$ and we found the corresponding veto parameters $E_f,\ E_c,\ r_c,\ t_f$ and $t_c$ which are functions of the rejection efficiencies.  

In this section we will assume that the muon tracking is perfect, leaving the general case to Sec.~\ref{trackcatt}.  Therefore we will only be interested in cylindrical vetoes.  The cylindrical veto is then fully described by a single parameter, the efficiency $\re_c$.  The showering energy threshold $E_c(\re_c)$, veto cylinder radius $r_c(\re_c)$ and veto duration $t_c(\re_c)$ may be read from Figs.~\ref{cilefig} and \ref{cilrt2fig} while the corresponding fractional dead time was reported in Fig.~\ref{ciltlfig}.

But what is the optimal value of $\re_c$?  This depends on the science goal considered.  For each science goal, we have calculated a statistic characterizing the goal.  This statistic depends on the total live time of the experiment and on the total background during that live time.  We have assumed that the experiment runs for 6 years, and so the live time will just be $6t_l$ years.  As $t_l$ is known as a function of $\re_c$ (Fig.~\ref{ciltlfig}), one may then obtain the statistic as a function of $\re_c$ alone and then choose the $\re_c$ which optimizes the statistic, optimizing the veto strategy for that science goal.  As the vetoes are not hardwired, there is no problem using a different veto strategy for each science goal.

\subsection{Neutrino mass hierarchy} \label{gersez}

\begin{figure} 
\begin{center}
\includegraphics[width=3.5in]{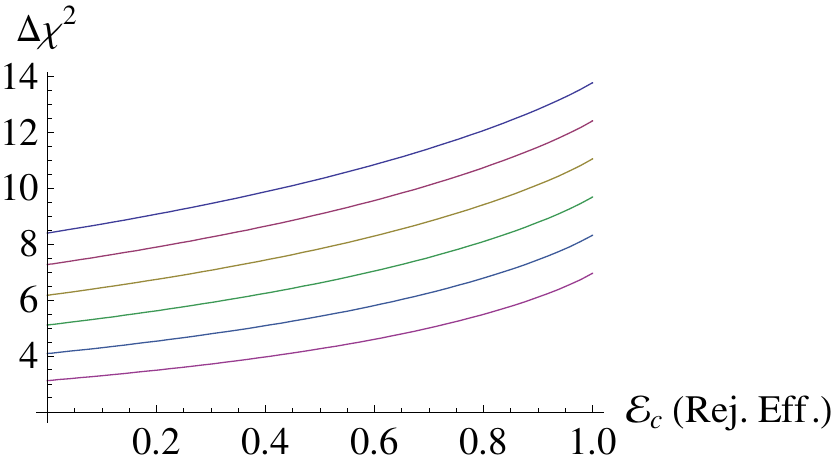}
\caption{The $\Delta\chi^2$ with which the neutrino mass hierarchy may be determined by JUNO in 6 years with a fraction $t_l$ of live time and ${}^9$Li rejection efficiency $\re_c$.  For the curves from top to bottom the fractions $t_l$ are 1.0, 0.9, 0.8, 0.7, 0.6 and 0.5.   It is assumed that the nonlinear energy response of the detector is known perfectly.}
\label{chifig}
\end{center}
\end{figure}

In this subsection we will optimize the veto strategy for the determination of the neutrino mass hierarchy at JUNO.  For this purpose, we have computed the statistic $\Delta\chi^2$ which is equal to the difference between the $\chi^2$ value of a best fit of a given spectrum to the normal and inverted hierarchies
\beq
\Delta\chi^2=\chi^2({\rm{normal}})-\chi^2({\rm{inverted}}). \label{dcdef}
\eeq
In Fig.~\ref{chifig} we display $\Delta\chi^2$ computed for an Asimov data set generated using the inverted hierarchy, using 6 years of run time, which throughout this paper we assume corresponds to $1.35\times 10^5$ inverse $\beta$ decay events.  

Note that $\Delta\chi^2$ is not that of Wilks' theorem, which is defined to be the difference between a best $\chi^2$ fit to a null hypothesis and that to a best fit of a continuous family of parameters of which the null hypothesis is one choice.  Indeed, the choice of hierarchies is not even continuous.  For example, the quantity $\Delta\chi^2$ defined in Eq.~(\ref{dcdef}) is Gaussian-distributed \cite{xinfiducia,fiducia} whereas that in Wilks' theorem is distributed following a $\chi^2$ distribution.  Therefore this $\Delta\chi^2$ is not the square of the number of $\sigma$ of confidence in a null hypothesis.  However, in a Bayesian framework in which one assigns a symmetric prior to the two hierarchies, this $\Delta\chi^2$ is simply related to the {\it{sensitivity}} $s$ to the hierarchy via the relation \cite{xinfiducia,fiducia}
\beq
s=\frac{1}{1+e^{-\Delta\chi^2/2}}.
\eeq
If this sensitivity is converted into a number of $\sigma$ using the error function, one finds a reduction of about $0.5\sigma$ as compared with the naive square root formula.

The detector energy resolution, number of events, etc. are as in Ref.~\cite{noisim} and conform to the experimental goals. However we assume that the nonlinear energy response of the detector is perfectly known.  In fact the experimental goal is a 1\% precision on the nonlinear energy response, which has a large effect on $\Delta\chi^2$ as has been foreseen in Ref.~\cite{xin2012} and demonstrated in Ref.~\cite {2rivelsim}.  However the precise effect on $\Delta\chi^2$ depends heavily on additional assumptions, such as the shape of the nonlinearity.  It may be possibile to estimate the nonlinearity by using the known general features of the spectrum \cite{juno}, however such a strategy may be ineffective or even problematic if the reactor neutrino spectrum has unexpected features, such as the 5 MeV bump first reported by RENO \cite{renobump} and later confirmed by Double Chooz \cite{doublebump} and Daya Bay \cite{dayabump}.  Therefore the inclusion of the unknown nonlinear energy response in this study would introduce uncertainties which would obfuscate the effect of the spallation backgrounds.

\begin{figure} 
\begin{center}
\includegraphics[width=3.5in]{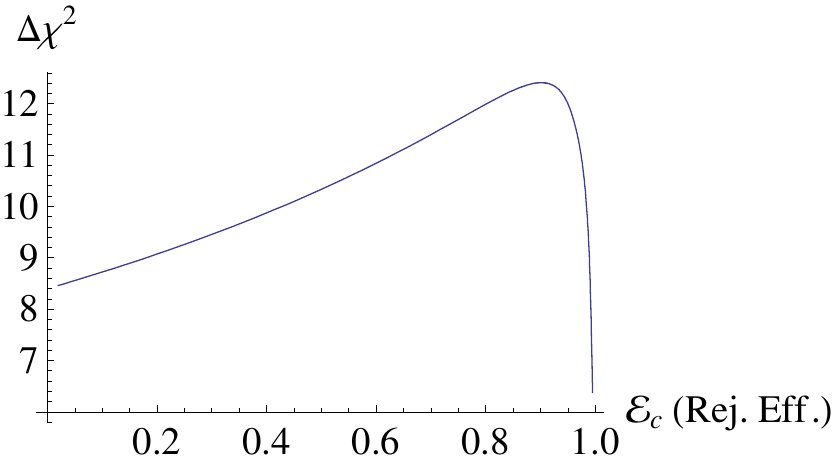}
\caption{The $\Delta\chi^2$ with which the neutrino mass hierarchy may be determined by JUNO in 6 years as a function of the rejection efficiency $\re_c$.  The optimal rejection efficiency is 90\%.}
\label{chi2fig}
\end{center}
\end{figure}

The optimization of the rejection efficiency $\re_c$ is now quite easy.  One need only substitute Fig.~\ref{ciltlfig}, which gives $t_l$ as a function of $\re_c$, into Fig.~\ref{chifig} which gives $\Delta\chi^2$ as a function of $t_l$ and $\re_c$ to obtain $\Delta\chi^2$ as a function of $\re_c$ alone, shown in Fig.~\ref{chi2fig}.  One can see that, in contrast with the common claim that the optimal rejection efficiency at JUNO will be at least 99\%, the optimal rejection efficiency is in fact only 90\% when the tracking is perfect.  Using Figs.~\ref{cilxfig}, \ref{cilefig}, \ref{cilrt2fig} and \ref{ciltlfig} with $\re_c=0.9$, one may obtain the corresponding optimal veto parameters, which are summarized in Table~\ref{cilparamtab}.


Despite the relatively large contamination, in the case of perfect tracking the ${}^9$Li background only reduces $\Delta\chi^2$ from 13.8 to 12.4.  This is our main result.

If Taishan reactors 3 and 4 are not built, then the reactor neutrino signal at JUNO will be reduced by about 26\%.  This can be incorporated into the present analysis with a 26\% reduction of the live time $t_l$ in Fig.~\ref{chifig}, resulting in a 20-25\% reduction in $\Delta\chi^2$.

\subsection{Uncertainty in the ${}^9$Li spectrum}

The small reduction in $\Delta\chi^2$ is a result of the fact that the shape of the ${}^9$Li decay background is known quite well, and so can be reliably subtracted.  What is the shape?  First, recall that we are only interested in decays producing neutrons
\beq
\rm {}^9Li\rightarrow {}^9 Be^* +e^-+\overline{\nu}_e\rightarrow n + 2\alpha+e^-+\overline{\nu}_e
\eeq
where ${}^9$Be${}^*$ is an excited state.  The $\overline{\nu}_e$ is not observed.  The spectrum of the $e^-$ is known quite precisely from Fermi's theory of $\beta$ decay.  It extends up to about 11 MeV and overlaps with the entire reactor neutrino prompt energy spectrum.  Indeed the scintillator detector cannot distinguish these $e^-$ from the $e^+$ produced by the inverse $\beta$ capture of reactor neutrinos.  The neutron and $\alpha$ particles are produced essentially immediately after the $\beta$ decay.  

However, as 90\% of neutron-emitting decays begin with a $\beta$ decay to the low energy (less than 3 MeV) excited states of ${}^9$Be \cite{isolde}, little energy is available for the neutron and $\alpha$ particles.  As a result, these particles have kinetic energies in general well below 2 MeV, often below 1 MeV.  In this regime their scintillator quenching factors are large.  The $\alpha$ particle quenching factor is more than 10 times that of the electron, reducing its contribution to the scintillation light by a factor of more than 10 so that its contribution to the observed energy negligible.  The neutron's quenching factor is more than 5 times that of the electron, so its contribution is small but not negligible.  The neutron's energy begins to arrive in the scintillator as it slows. The neutrons slows as it collides with particles in the scintillator.  The time between such collisions is about 5 ns and about half of the collisions, those with free protons, reduce the neutron's energy by about one half.  Therefore the neutron loses most of its energy in about 10 ns.  

While one might think that the PMTs could then distinguish the neutron's energy from that of the electrons using the 10 ns time difference and so veto the ${}^9$Li events, this is not the case.  The main problem is the time-lag before the scintillator scintillates, which ranges from 10s to 100s of ns.  Another problem is that JUNO is about the same size as the mean free path of the ultraviolet photons created by the wave shifters.  As a result, a significant portion of the photons created by the electron bounce inside of the scintillator before arriving at the PMTs, and so arrive 10s of nanoseconds late.  These late photons from the electron cannot be distinguished from the photons from the neutron slowing, and so one only observes the total energy.  However, the neutron energy was already 5-10 times smaller than the electron energy and the quenching factor reduces its contribution to the visible energy by an additional factor of 5, thus the neutron contributes only about 5\% of the visible energy.  Now the neutron spectrum, although it has been measured by many experiments \cite{cc1,cc2,brown}, is not known as precisely as the electron spectrum.  Nonetheless, the fact that it contributes only 5\% of the visible energy means that the uncertainty in the neutron spectrum yields a very small uncertainty in the ${}^9$Li spectrum.  This is why the ${}^9$Li spectrum is so well known, which allows the presence of a large ${}^9$Li contaminant in the IBD event sample to have such a small effect on the sensitivity to the hierarchy.  JUNO itself will provide a very precise measurement of the quenching factor weighted spectrum, actually of a weighted combination of the ${}^9$Li and ${}^8$He spectra, which will only decrease this uncertainty.

In fact, the biggest uncertainty in the ${}^9$Li spectrum may arise from the veto efficiency $\re_c$, which depends upon the the precision of the reconstruction of the muon track and the radial distribution $f(R)$ of the ${}^9$Li which is not well known, although JUNO will measure it better than it has ever been measured before.  Unlike the previous uncertainty, this does not affect the shape of the observed energy spectrum, but only the overall normalization.  The precision of these estimates are quite difficult to forecast.  Therefore in this note we simply assume that the shape of the ${}^9$Li spectrum is known perfectly and we include a 1\% uncertainty in its overall normalization.


\subsection{Mixing angle $\theta_{12}$}

In this subsection we apply the same analysis to the determination of $\theta_{12}$ as that used for the determination of the hierarchy in Subsec.~\ref{gersez}.  We will choose the veto strategy to minimize $\sigma(\spp2212)$, the precision with which $\spp2212$ can be determined.   There are various calculations of this uncertainty in the literature.  The earliest \cite{petcov12,kaoru} do not consider the uncertainty in the shape of the reactor neutrino spectrum.  Later calculations, such as Ref.~\cite{whitepaper}, do consider this uncertainty but use the theoretical lower bounds on the uncertainty given in Ref.~\cite{huber}.  However, it is now known that the lower bound on this uncertainty is far from saturated, indeed it would imply that the 5 MeV bump~\cite{renobump,dayabump} is excluded at $4\sigma$.

\begin{figure} 
\begin{center}
\includegraphics[width=3.5in]{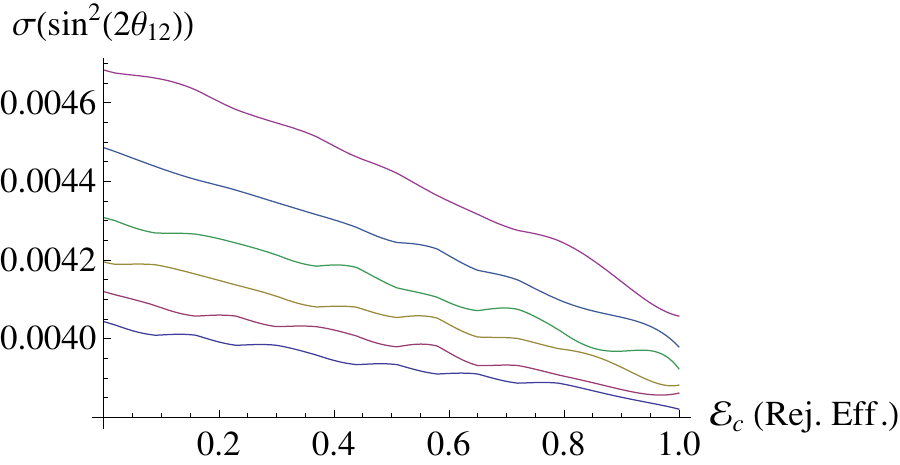}
\caption{The $1\sigma$ uncertainty with which $\spp2212$ can be measured at JUNO in 6 years with a fraction $t_l$ of live time and ${}^9$Li rejection efficiency $\re_c$, from Ref.~\cite{noi12}.  For the curves from bottom to top the fractions $t_l$ are 1.0, 0.9, 0.8, 0.7, 0.6 and 0.5.   The main contribution to the uncertainty arises from the uncertainty in the reactor spectrum measured by Daya Bay \cite{dayabump}.}
\label{sig12fig}
\end{center}
\end{figure}

This motivates us to use the more recent determination of $\sigma(\spp2212)$ from Ref.~\cite{noi12}, shown in Fig.~\ref{sig12fig}, which uses the observational uncertainty on the reactor neutrino flux from Daya Bay \cite{dayabump}.  In addition, this is the only study which yields  $\sigma(\spp2212)$ as a function of the dead time and the ${}^9$Li veto efficiency, which is necessary to optimize the vetoes.  

The most obvious distinction between Fig.~\ref{sig12fig} and the corresponding figure for the mass hierarchy, Fig.~\ref{chifig}, is that in the former the experimental performance hardly depends upon the rejection efficiency and the dead time in the range plotted.  This is because the precision of a determination of $\theta_{12}$ is dominated by systematic errors, the uncertainty in the reactor spectrum, unlike the determination of the hierarchy.  Of course, if we included the uncertainty in the detector's energy response then then the systematic errors would play a more important role in the determination of the hierarchy.  However, so long as this response can be understood well enough to achieve a reasonable determination of the hierarchy, it will play a subdominant role.

\begin{figure} 
\begin{center}
\includegraphics[width=3.5in]{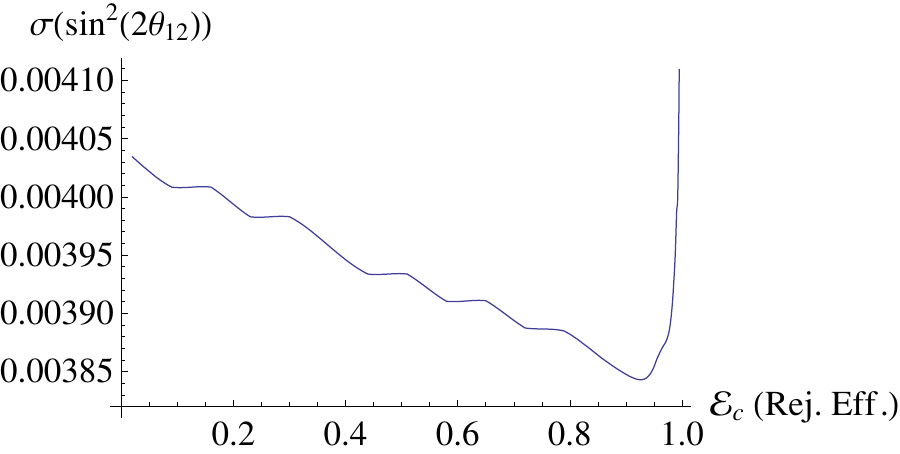}
\caption{The  $1\sigma$ uncertainty with which $\spp2212$ can be measured by JUNO in 6 years as a function of the rejection efficiency $\re_c$.  The optimal rejection efficiency is 93\%, higher than for the hierarchy because the uncertainty is dominated by a systematic error, the uncertainty in the reactor neutrino spectrum, which is taken to be as in Fig.~\ref{sig12fig}.}
\label{sig122fig}
\end{center}
\end{figure}

As in the case of the hierarchy, we now substitute Fig.~\ref{ciltlfig}, which gives $t_l$ as a function of $\re_c$, into Fig.~\ref{sig12fig} which gives $\sigma(\spp2212)$ as a function of $t_l$ and $\re_c$ to obtain $\sigma(\spp2212)$ as a function of $\re_c$ alone, shown in Fig.~\ref{sig122fig}.   Note that the optimal rejection efficiency is now 93\%, as compared with 90\% for the mass hierarchy and the precision $\sigma(\spp2212)$ is $0.0038$ or $0.45\%$, as compared with $0.44\%$ with no background at all.  As summarized in Table~\ref{cilparamtab} this means that the optimal veto strategy for the determination of $\theta_{12}$ should have 6\% dead time, as compared with only 3\% for the neutrino mass hierarchy.  The reason for this is that the error in $\theta_{12}$ is dominated by the systematic error in the reactor spectrum, not by statistical fluctuations like the measurement of the hierarchy.  Therefore a larger dead time in the $\theta_{12}$ analysis has less effect on the precision of the measurement than it would in a determination of the hierarchy.  

Physically this is a consequence of the fact that $\theta_{12}$ is determined by a broad feature in the reactor neutrino spectrum, essentially the difference between the neutrino flux at intermediate energies at the solar oscillation maximum and that at higher and lower energies.  Being a broad feature it is insensitive to the energy resolution of the detector and more importantly it involves essentially the whole sample of neutrinos.   Thus the statistical uncertainty on $\theta_{12}$ is extremely small.  The determination of the hierarchy is quite different.  It comes from comparing $\Delta M^2_{ee}$ \cite{parke2005} measured at intermediate and high energies with the position of the 15th, 16th and 17th peaks at low energies \cite{noiteor,kaoru}.  These low energy peaks contain few events, especially when one considers the effect of the finite energy resolution and interference from neutrinos coming from reactors at distinct baslines \cite{noiteor} at JUNO.  Therefore statistical uncertainty is very important in the determination of the hierarchy, which drives the preference for a veto strategy with little dead time.

\begin{table}[position specifier]
\centering
\begin{tabular}{c|l|l|l|l|l|l|l}
Science Goal&Statistic&$\re_c$&$t_l$&$E_c$&$x$&$t_c$&$r_c$\\
\hline\hline
Mass hierarchy&$\Delta\chi^2=12.4$&90\%&97\%&2.9 GeV&95\%&1.1 sec&2.3 m\\
\hline
Measure $\theta_{12}$&$\frac{\sigma(\spp2212)}{\spp2212}=0.45\%$&93\%&94\%&2.4 GeV&96\%&1.2 sec&2.5 m\\
\hline
\end{tabular}
\caption{The optimal cylindrical veto parameters for the determination of the mass hierarchy and $\theta_{12}$ assuming perfect tracking}
\label{cilparamtab}
\end{table}

\section{Imperfect Tracking} \label{trackcatt}

In this section we will consider the effects of imperfect tracking of muons.  We will make the poor approximation that the probability of tracking a muon is independent of its showering energy.  Up to this point the generalization to muon bundles has been straightforward, it only required modifying the functions $m(E)$ and $l(E)$ by including the weighted sum of bundles.  However if one introduces bundles, then one either needs to introduce new parameters describing how well the bundles are tracked, or else make the further crude approximation that the tracking of each muon is independent of whether it arrives in a single or multimuon event.  In the latter case, the treatment below remains applicable, one need only remember that $\alpha$ and $\beta$, which will be defined momentarily, are fractions or muons, not of muon bundles.

\begin{table}[position specifier]
\centering
\begin{tabular}{c|l}
Variable&Tracking\\
\hline\hline
$1-\alpha-\beta$&Muon well tracked\\
\hline
$\alpha$&Muon poorly tracked but believed to be well tracked\\
\hline
$\beta$&Muon poorly tracked and known to be poorly tracked\\
\hline
\end{tabular}
\caption{The tracking of muons is characterized by the parameters $\alpha$ and $\beta$, assumed to be independent of the showering energy}
\label{abtab}
\end{table}

With these caveats, the tracking is described by just two parameters, $\alpha$ and $\beta$, which, as is summarized in Table~\ref{abtab} are the fraction of muons for which the tracking fails but is believed to succeed and the fraction of muons for which the tracking fails but the experimenter realizes that the tracking has failed.  Our veto strategy then demands that full detector vetoes be applied to a fraction $\beta$ of all muons.  In the case of the $\alpha$ muons a cylindrical veto is applied, but as it is quite unlikely that this cylinder coincides accidentally with the muon track, we will simply assume that the veto does not increase the rejection efficiency.  The fraction of muons for which tracking is successful is $1-\alpha-\beta$.

For simplicity we will make the crude approximation that the tracks of well-tracked muons are determined precisely.  In fact, one expects a precision of only 1-2 meters.  However, this precision depends on the configuration of the top veto, which is not yet fixed, and so cannot be estimated reliably at this time.  The result of the uncertainty in the muon track position would be to increase the optimal $r_c$ by 1 meter or more and indeed it may be worthwhile to consider a cylinder which gets wider along the muon's trajectory.    As a consequence, the total dead time will be higher than that estimated below and the rejection efficiency lower.

The resulting fractional live time is
\beq
t_l(E_c,r_c,x)={\rm{exp}}\left(-(1-\beta)\frac{\pi d(E_c) r_c^2}{V} t_c m(E_c)-\beta t_f m(E_f) \right). \label{vivototale}
\eeq
The veto strategy now depends on two parameters, the rejection efficiencies $\re_f$ and $\re_c$ for the full detector and the cylindrical vetoes.  On the other hand, the science goals only depend upon $t_l$ and the total rejection efficiency
\beq
\re=(1-\alpha-\beta)\re_c+\beta\re_f. \label{duere}
\eeq
Therefore the optimization of the veto strategy will consist of two steps.  First, for each $\re$, we will find $\re_c$ and $\re_f$ so as to maximize $t_l$.  Next, in the following two subsections, we will choose $\re$ for each science goal so that the corresponding statistic is optimized.

The total live time $t_l$ in Eq.~(\ref{vivototale}) has one very useful feature.  It is monotonic in both the live times $t_l$ in Eq.~(\ref{pienot}) and in Eq.~(\ref{ciltl}) of the full detector and cylindrical vetoes, in fact it is just a product of a positive powers of the two live times
\beq
t_l=\left[t_l {\rm{\ (full\ detector)}}\right]^{\beta}\left[t_l {\rm{\ (cylindrical)}}\right]^{1-\beta}.  
\eeq
Therefore the $t_f(\re_f)$, $E_f(\re_f)$, $t_c(\re_c)$, $r_c(\re_c)$ and $E_c(\re_c)$ that minimized the $t_l$ of Eqs.~(\ref{pienot}) and (\ref{ciltl}) for each individual veto in Sec.~\ref{taglisez} also minimize the total $t_l$ of Eq.~(\ref{vivototale}).   This greatly simplifies our analysis, as it means that we can continue to use the functions $t_f(\re_f)$, $E_f(\re_f)$, $t_c(\re_c)$, $r_c(\re_c)$ and $E_c(\re_c)$ that we found assuming perfect tracking.

\begin{figure} 
\begin{center}
\includegraphics[width=3in]{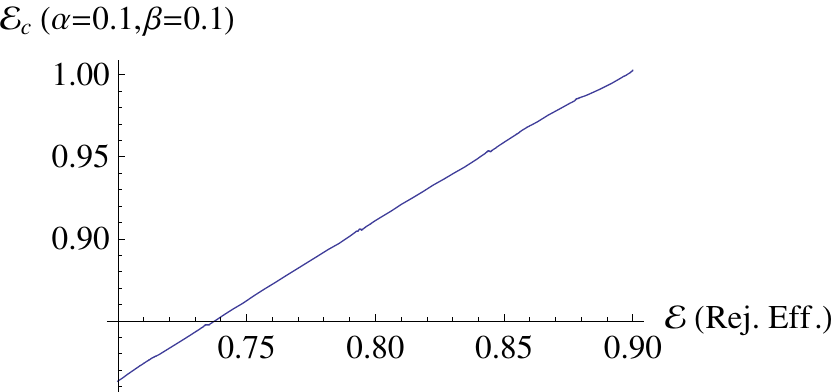}
\includegraphics[width=3in]{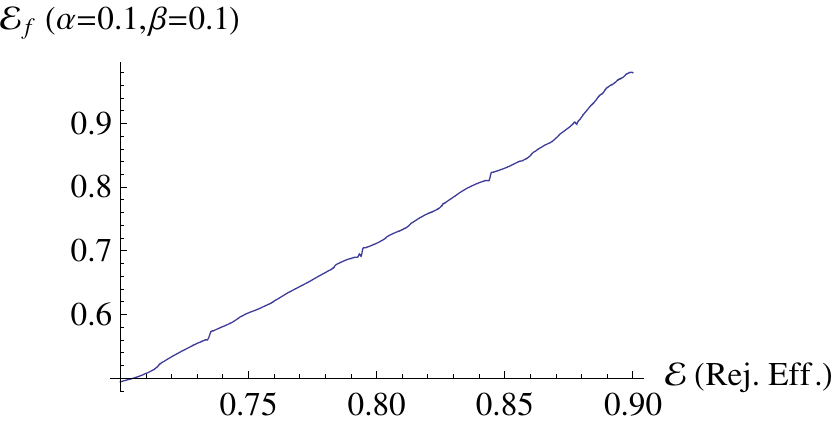}
\caption{The optimal cylindrical and full detector veto efficiencies $\re_c$ and $\re_f$ as a function of the total veto efficiency $\re$.  Here $\alpha=\beta=0.1$ which means that 80\% of muons are well tracked but 90\% are believed to be well tracked.}
\label{recreffig}
\end{center}
\end{figure}

Our goal in this section is to obtain $\re_f(\re)$ and $\re_c(\re)$.  When $\beta\neq 0$, the former is easily obtained from Eq.~(\ref{duere})
\beq
\re_f(\re)=\frac{\re-(1-\alpha-\beta)\re_c(\re)}{\beta}. \label{ref}
\eeq
Thus we need only substitute $\re_f(\re)$ from Eq.~(\ref{ref}) into the expression~(\ref{vivototale}) for the total live time and then choose $\re_c(\re)$ so as to minimize the total live time.  

However the function $\re_c(\re)$ also depends on the parameters $\alpha$ and $\beta$ and so it is not easy to display in a figure.  While we have done this optimization for every $\alpha$ and $\beta$, in this subsection we display, in Fig.~\ref{recreffig}, the optimal cylindrical and full detector veto efficiencies only for $\alpha=\beta=0.1$.  This corresponds to the case in which the tracking is successful for 80\% of muons but the experimenter believes that it has been successful for 90\% of muons, therefore the cylindrical veto is used 90\% of the time.  Note that in general the optimal full detector rejection efficiency is lower than the rejection efficiency for the cylindrical vetoes.  This is because full detector vetoes lead to a much larger dead time.  The maximum possible rejection efficiency $\re$ is $1-\alpha=90\%$, as a fraction $\alpha=10\%$ of the muons are poorly tracked unbeknownst to the experimentalist and so the cylindrical veto covers the wrong region.

\begin{figure} 
\begin{center}
\includegraphics[width=3.5in]{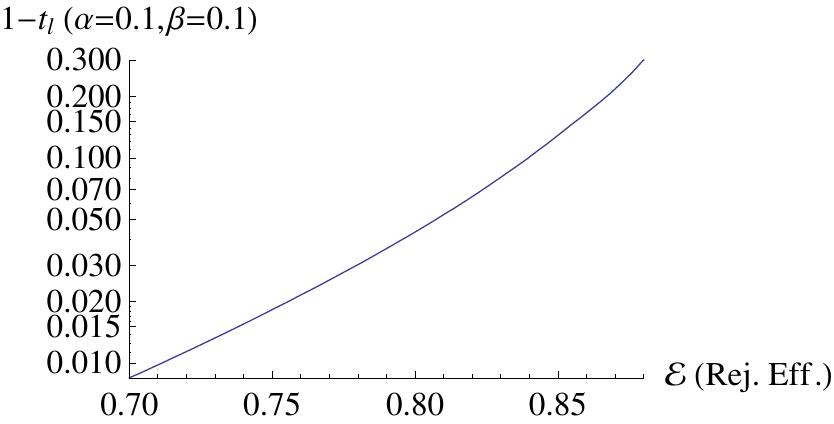}
\caption{The fractional dead time which is required to obtain a total rejection efficiency $\re$ with $\alpha=\beta=0.1$}
\label{retlfig}
\end{center}
\end{figure}

The corresponding fractional dead time $1-t_l$ is plotted in Fig.~\ref{retlfig}.  Although only 80\% of the muons are successfully tracked, the fact that half of the poorly tracked muons are known to be poorly tracked nonetheless allows a rejection efficiency of 85\% with a dead time of 13\%, comparable to KamLAND.

\subsection{The neutrino mass hierarchy}

\begin{figure} 
\begin{center}
\includegraphics[width=3.5in]{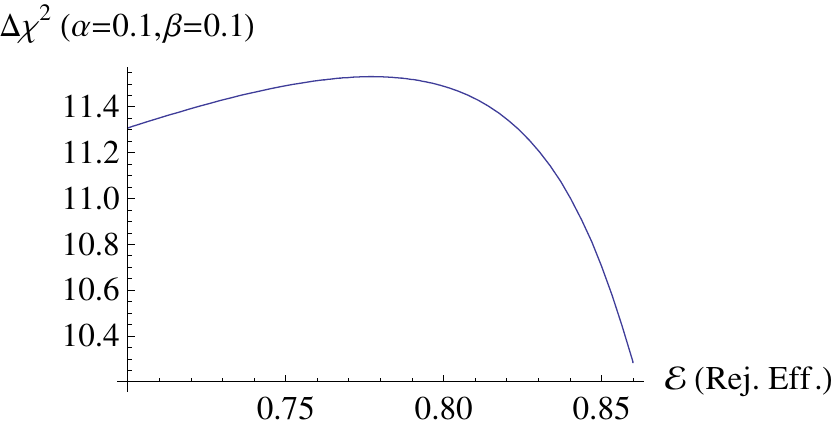}
\caption{The statistic $\Delta\chi^2$ which characterizes the sensitivity to the neutrino mass hierarchy as a function of the total rejection efficiency $\re$ with $\alpha=\beta=0.1$}
\label{rechifig}
\end{center}
\end{figure}

The optimization of the veto for the neutrino mass hierarchy is now quite simple.  Only one independent parameter remains to be fixed, the total rejection efficiency $\re$.  Similarly to the case with perfect tracking, the sensitivity to the hierarchy as a function of $\re$ is obtained by restricting Fig.~\ref{chifig} to the pairs $(\re,t_l)$ on the curve in Fig.~\ref{retlfig}, corresponding to the maximal live time achievable for each rejection efficiency.  This yields Fig.~\ref{rechifig}.  Here one can see that with $\alpha=\beta=0.1$, the optimal rejection efficiency is 78\% and it yields a $\Delta\chi^2$ of 11.5, as compared with 13.8 with no background and 12.4 with perfect tracking. The corresponding optimal veto parameters are summarized in Table~\ref{totparamtab}.

\begin{figure} 
\begin{center}
\includegraphics[width=3in]{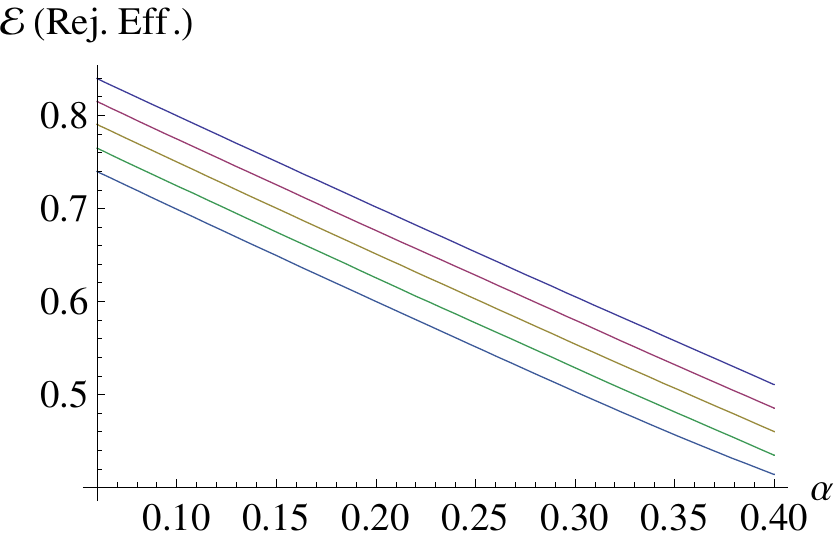}
\includegraphics[width=3in]{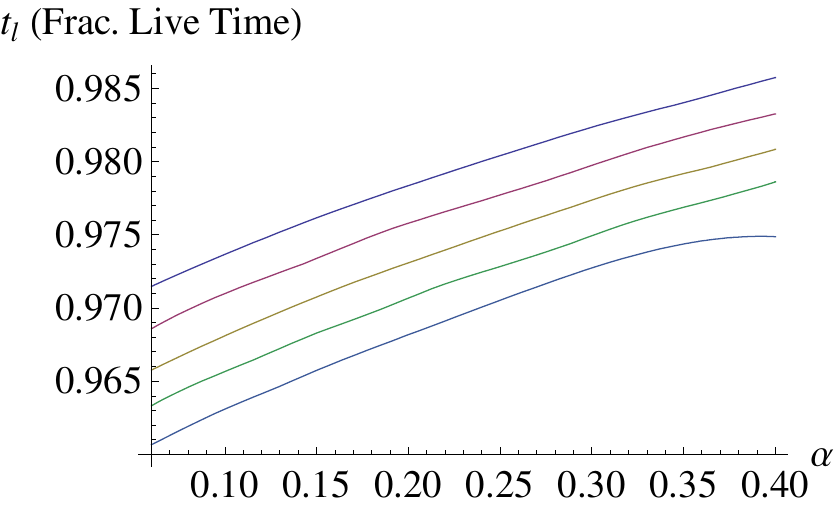}
\caption{The optimal veto efficiency and total fractional live time for the determination of the mass hierarchy as a function of the tracking efficiency parameters $\alpha$ and $\beta$.  The curves, from top to bottom, correspond to $\beta=0,\ 0.1,\ 0.2,\ 0.3$ and $0.4$.}
\label{efftlfig}
\end{center}
\end{figure}

\begin{figure} 
\begin{center}
\includegraphics[width=3.5in]{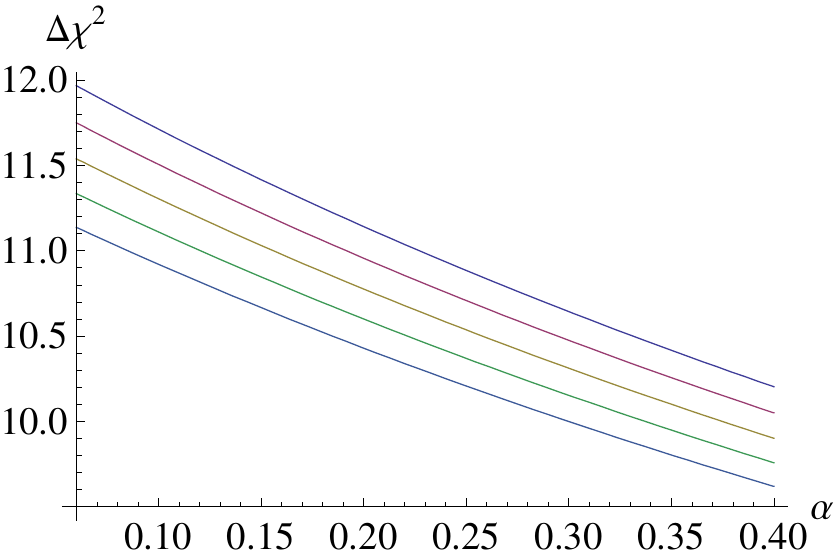}
\caption{The statistic $\Delta\chi^2$ which characterizes the sensitivity to the neutrino mass hierarchy as a function of the fraction $\beta$ of muons which are known to be poorly tracked and the fraction $\alpha$ of muons which appear to be well tracked but in fact are poorly tracked. The curves, from top to bottom, correspond to $\beta=0,\ 0.1,\ 0.2,\ 0.3$ and $0.4$.}
\label{chifinalefig}
\end{center}
\end{figure}

For general tracking efficiency parameters $\alpha$ and $\beta$, one can follow this same strategy to optimize the total rejection efficiency $\re$.  The results are shown in Fig.~\ref{efftlfig} together with the corresponding fractional live times obtained using Fig.~\ref{retlfig}. The corresponding statistic $\Delta\chi^2$ is presented in Fig.~\ref{chifinalefig}.  This is one of our main results.  It shows the dependence of the sensitivity to the hierarchy on the tracking.  It is evident that $\alpha$ has a greater impact upon $\Delta\chi^2$ than $\beta$, reflecting the fact that if it is known that the tracking has failed (as occurs for a fraction $\beta$ of muons) then one can nonetheless perform a full detector veto which increases the sensitivity to the hierarchy.  One can see that the failure to track $10\%$ of muons does not greatly reduce the sensitivity to the hierarchy.  However, muon bundles are responsible for about 30\% of muon events, and so the ability to track bundles does have a nonnegligible effect on the sensitivity.  KamLAND has several classifications of how reliably a given muon was tracked \cite{lindley}.  A tighter reliability threshold corresponds to a larger value of $\beta$ and a smaller value of $\alpha$.  The results of our study can be used to optimize this reliability threshold.

\begin{figure} 
\begin{center}
\includegraphics[width=3in]{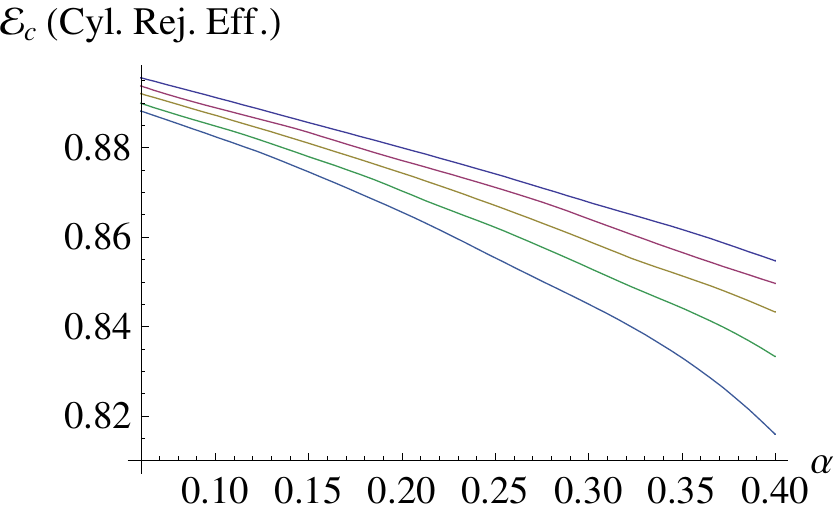}
\includegraphics[width=3in]{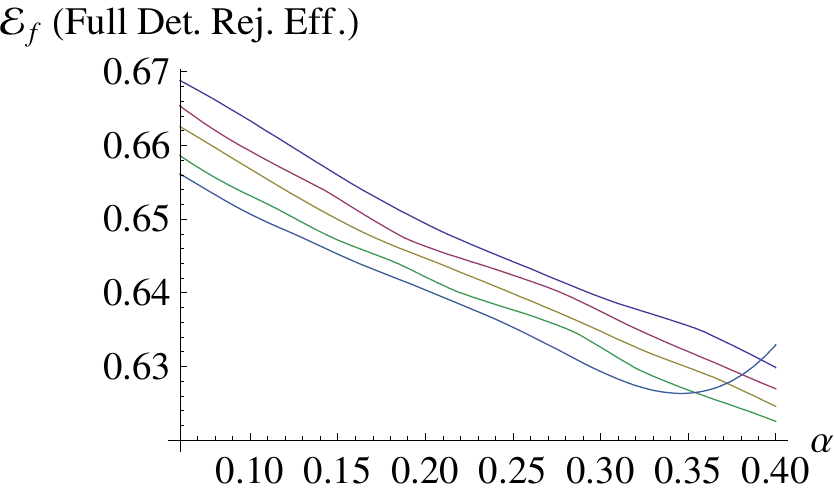}
\caption{The optimal cylindrical and full detector veto efficiencies as a function of the tracking efficiency parameters $\alpha$ and $\beta$.  The curves, from top to bottom, correspond to $\beta=0,\ 0.1,\ 0.2,\ 0.3$ and $0.4$.}
\label{refinalefig}
\end{center}
\end{figure}

\begin{figure} 
\begin{center}
\includegraphics[width=2.1in]{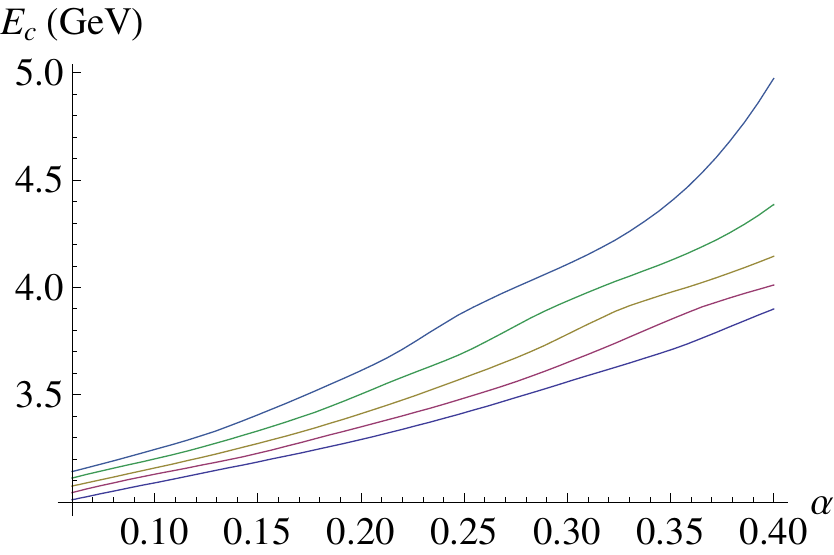}
\includegraphics[width=2.1in]{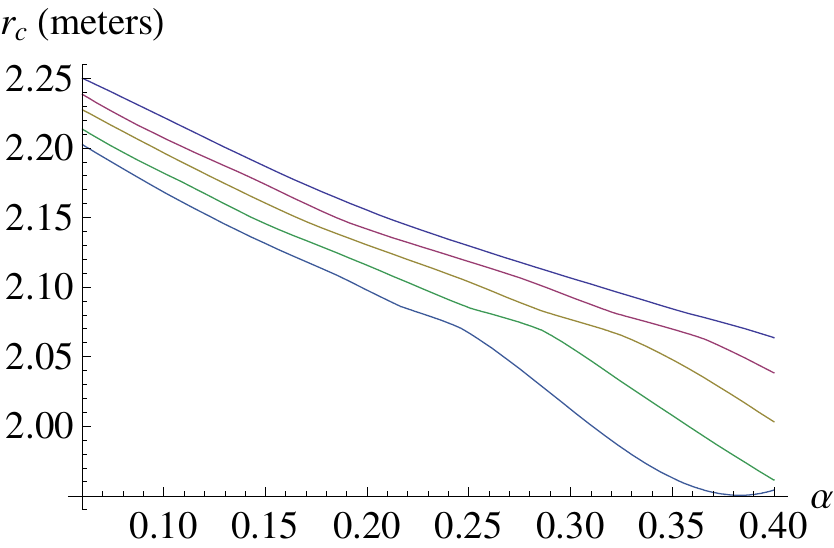}
\includegraphics[width=2.1in]{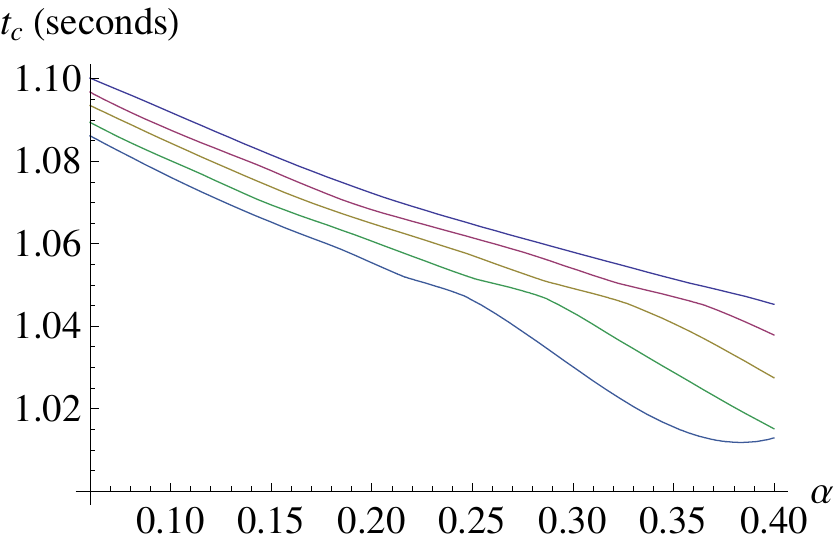}
\caption{The optimal cylindrical veto parameters as a function of the tracking efficiency parameters $\alpha$ and $\beta$.  The curves, from top to bottom, correspond to $\beta=0,\ 0.1,\ 0.2,\ 0.3$ and $0.4$ except for the left panel where the order is inverted}
\label{ercfinalefig}
\end{center}
\end{figure}

\begin{figure} 
\begin{center}
\includegraphics[width=3in]{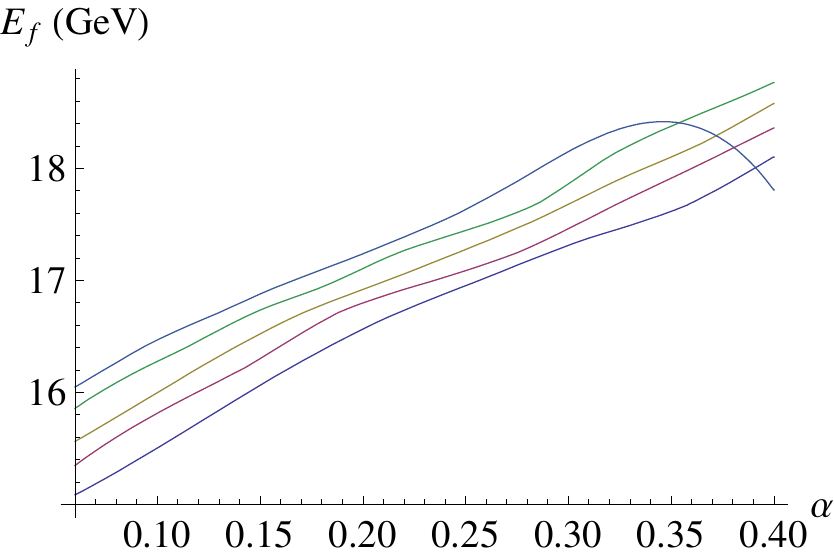}
\includegraphics[width=3in]{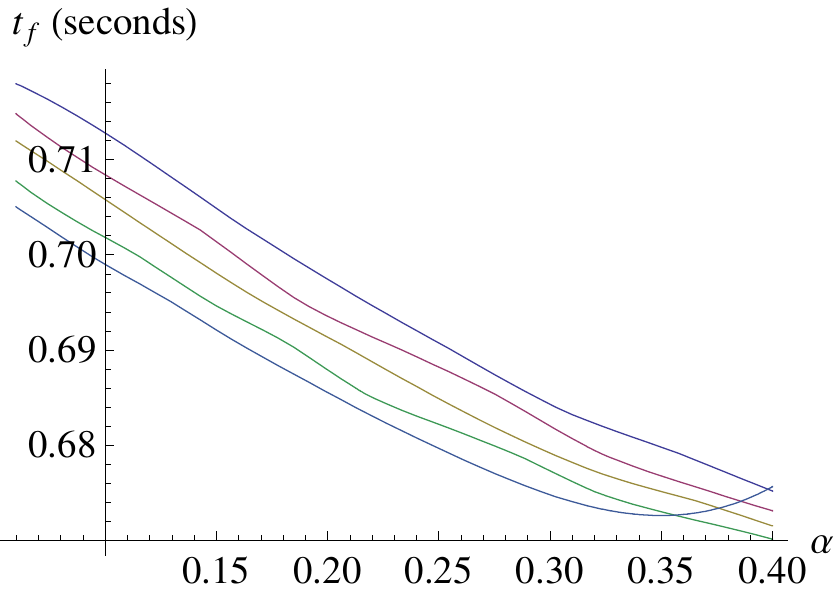}
\caption{The optimal full detector veto parameters as a function of the tracking efficiency parameters $\alpha$ and $\beta$.  The curves in the left (right) panel from bottom to top (top to bottom) correspond to $\beta=0,\ 0.1,\ 0.2,\ 0.3$ and $0.4$}
\label{ecfinalefig}
\end{center}
\end{figure}

We have now optimized the veto strategy, for the determination of the neutrino mass hierarchy, as a function of $\alpha$ and $\beta$.  For completeness, we will now describe this optimal veto.  In Fig.~\ref{refinalefig} one can find the individual veto efficiencies for the cylindrical and full detector vetoes.  Surprisingly, the individual efficiencies are quite independent of the tracking, one always wants a rejection efficiency in the mid to high eighties for the cylindrical veto and in the mid  sixties for the full detector veto.  As these individual efficiencies give the veto parameters, using the results of Sec.~\ref{taglisez}, this implies that the optimal veto parameters themselves are quite insensitive to the tracking abilities of the detector.  The optimal parameters are shown in Figs.~\ref{ercfinalefig} and~\ref{ecfinalefig} for the cylindrical and full detector vetoes respectively.

\subsection{Measuring $\theta_{12}$}

\begin{figure} 
\begin{center}
\includegraphics[width=3.5in]{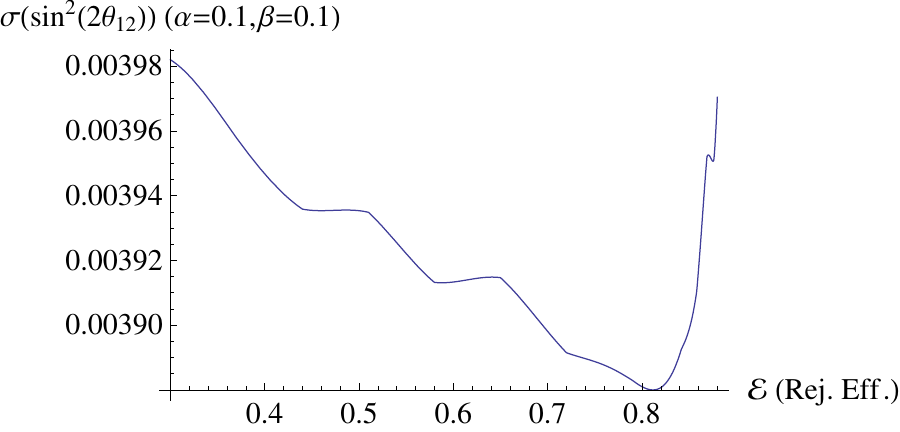}
\caption{The precision with which $\spp2212$ can be measured as a function of the total rejection efficiency $\re$ with $\alpha=\beta=0.1$}
\label{resigfig}
\end{center}
\end{figure}

The measurement of $\theta_{12}$ can be analyzed similarly to the hierarchy determination.  The precision $\sigma(\spp2212)$ as a function of $\re$ is obtained by restricting Fig.~\ref{sig12fig} to the pairs $(\re,t_l)$ on the curve in Fig.~\ref{retlfig}, corresponding to the maximal live time achievable for each rejection efficiency.  This yields Fig.~\ref{resigfig}.  Here one can see that with $\alpha=\beta=0.1$, the optimal rejection efficiency is 81\% and it yields a $\sigma(\spp2212)$ of $0.0039$ or $0.46\%$, as compared with $0.44\%$ with no background or $0.45\%$ with perfect tracking. The corresponding optimal veto parameters are summarized in Table~\ref{totparamtab}.

\begin{figure} 
\begin{center}
\includegraphics[width=3.5in]{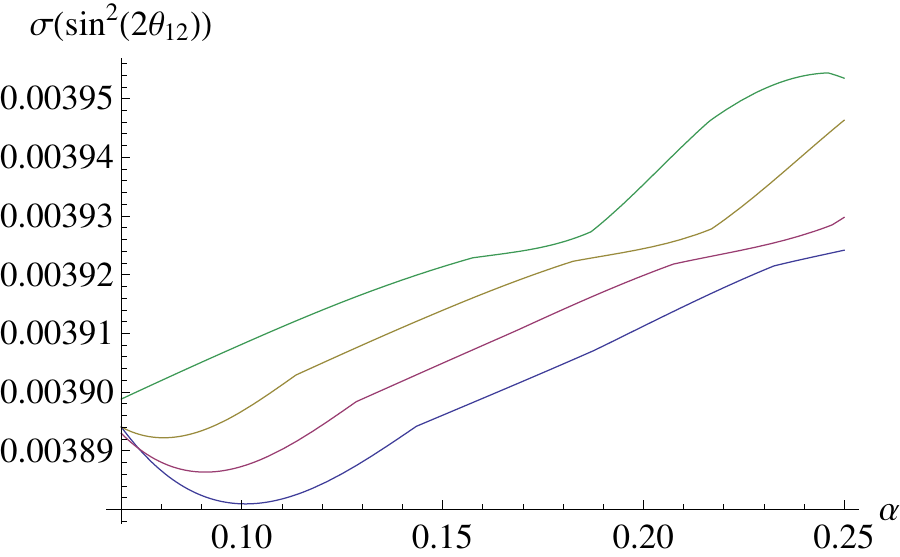}
\caption{The  precision with which $\spp2212$ can be measured as a function of the fraction $\beta$ of muons which are known to be poorly tracked and the fraction $\alpha$ of muons which appear to be well tracked but in fact are poorly tracked. The curves, from bottom to top, correspond to $\beta=0.1,\ 0.2,\ 0.3$ and $0.4$. For $\alpha>0.25$ our maximization did not converge sufficiently quickly to be displayed here.}
\label{sigfinalefig}
\end{center}
\end{figure}

For general tracking efficiency parameters $\alpha$ and $\beta$, one can follow this same strategy to optimize the veto strategy.   The corresponding statistic $\sigma(\spp2212)$ is presented in Fig.~\ref{sigfinalefig}.  One sees that the precision with which $\theta_{12}$ can be determined is essentially independent of the tracking.  By performing the same analysis as in the case of the hierarchy, we have also found that the optimal veto strategy is also quite insensitive to the tracking, and so for brevity we will not repeat the optimal strategy here in general.   In the case $\alpha=\beta=0.1$ the optimal strategy is summarized in Table \ref{totparamtab}.

\begin{table}[position specifier]
\centering
\begin{tabular}{c|l|l|l|l|l|l|l|l|l|l}
Goal&Stat&$\re$&$t_l$&$\re_c$&$\re_f$&$E_c$&$t_c$&$r_c$&$E_f$&$t_f$\\
\hline\hline
MH&$11.5$&78\%&97\%&89\%&66\%&3.1 GeV&1.1 s&2.2 m&16 GeV&0.7 s\\
\hline
$\spp2212$&0.46\%&81\%&95\%&92\%&73\%&2.5 GeV&1.1 s&2.3 m&12 GeV&0.8 s\\
\end{tabular}
\caption{The optimal cylindrical veto parameters for the determination of the mass hierarchy and $\theta_{12}$ assuming tracking parameters $\alpha=\beta=0.1$ corresponding to 20\% of muons being poorly tracked of which half are known to be poorly tracked.  The statistics reported are as in Table~\ref{cilparamtab}.}
\label{totparamtab}
\end{table}

\section{Conclusions}
The next generation of medium baseline neutrino experiments have two main goals, the determination of the neutrino mass hierarchy and a precise measurement of $\theta_{12}$.  The challenges involved in these two goals are different.  The first requires a precise measurement of the energies of the 15th,  16th and 17th $1-3$ oscillation peaks in the reactor neutrino spectrum \cite{noiteor}.  These are small and contain few events, therefore statistical fluctuations will be quite important.  The second instead requires a  measurement of the depth of the solar oscillation maximum, a broad feature which extends across the spectrum visible at these experiments.  This feature includes essentially all of the events in the sample, and so it will be measured extremely precisely at these experiments, with an uncertainty which is dominated by a systematic uncertainty in the reactor neutrino spectrum \cite{noi12}.

As statistical fluctuations play a central role in the determination of the hierarchy but not in the measurement of $\theta_{12}$, the optimal ${}^9$Li veto strategies are quite different for these two science goals.  The first is optimized with a very loose veto, which even with perfect tracking has an efficiency of only $90\%$.  Of course, given the high rate of muon bundles and showering muons \cite{noimuoni}, the tracking will not be perfect and so the efficiency will be appreciably lower.  On the other hand, as statistical fluctuations play a secondary role in the measurement of $\theta_{12}$, a tighter veto is optimal, with an efficiency as high as 93\% in the idealized case of perfect tracking.  This is not a result of a large impact of the spallation background on the precision of the measurement of $\theta_{12}$.  Indeed, with the 1\% uncertainty assumed here in the spallation background rate, even with no veto at all the precision in the measurement of $\spp2212$ would only increase from $0.44\%$ to $0.47\%$.  On the contrary, it is a result of the fact that a 6 year run is more than is necessary for a precise measurement of $\theta_{12}$, and so additional dead time has little impact on the precision.  

Indeed, a smaller detector with a less PMT coverage could measure $\theta_{12}$ nearly as well.  More to the point, given the central role of systematic errors, we suspect \cite{2rivelsim} that the most precise measurement of $\theta_{12}$ at the same price would be given by two smaller detectors, with low PMT coverage, at sufficiently distinct baselines so as to break the degeneracy between the unknown reactor spectrum and $\theta_{12}$.  Even if this would preclude a determination of the hierarchy, it may be the option for RENO 50 which would maximize synergy with JUNO and open the door to the measurement of the leptonic CP-violating phase in the years beyond~\cite{noidarts}.

\section* {Acknowledgement}
\noindent
JE and EC are supported by NSFC grant 11375201.   XZ is supported in part by  NSFC grants 11121092, 11033005 and 11375202.   EC  is also supported by the Chinese Academy of Sciences President's International Fellowship Initiative grant 2015PM063.  MG  is supported by the Chinese Academy of Sciences President's International Fellowship Initiative grant 2015PM007.


\end{document}

In the next decade, the experiments JUNO \cite{juno} and RENO 50 \cite{reno50} and perhaps LENA \cite{lena} will employ the largest liquid scintillator detectors ever constructed to detect antineutrinos from a variety of sources.  While these detectors will be underground to shield them from cosmogenic muons, the first two will be at the relatively modest depths of 700 and 900 meters respectively.  The showers induced by some of these muons will react with the carbon in the liquid scintillator and create ${}^9$Li and ${}^8$He some of whose decays yield the same double coincidence \cite{cr} used to identify antineutrinos via inverse beta decay (IBD).  In previous experiments, such as KamLAND, the resulting background was largely eliminated by vetoing events occurring soon after the passage of cosmogenic muons \cite{kamlandback}.  However in the case of these two new, large detectors such a veto is delicate as the time between muon events is comparable to both the lifetimes of ${}^9$Li and of ${}^8$He.  In this paper we present the results of a series of simulations of muons in such detectors.  The results of these simulations on the one hand indicate that KamLAND's cuts cannot be straightforwardly applied to JUNO and RENO 50 but on the other hand can be used to determine the efficiency of any new veto scheme.

We used the FLUKA simulation package \cite{fluka} to simulate the propagation of cosmogenic muons and antimuons inside of a 20 kton spherical detector consisting of a LAB-based liquid scintillator.  This corresponds to the detector proposed for the experiment JUNO, while the RENO 50 detector will be 18 kton and cylindrical.  We considered several different muon spectra, corresponding to various overburdens, topographies and cosmogenic muon distribution models.  We will report results obtained using the cosmogenic muon distribution of Ref.~\cite{bundle} which is illustrated in Fig.~\ref{initfig}.   As $\mu^-$ and $\mu^+$ interactions in the detector result in different isotope production rates \cite{lindley}, the ratio of $\mu^+$ to $\mu^-$ will affect our results.  We will assume an energy independent ratio of 1.37 $\mu^+$ per every $\mu^-$ corresponding to the value measured  by Kamiokande for 1.2 TeV muons in Ref.~\cite{antimu}.  In Sec.~\ref{docciasez} we will describe the distribution of the muon's energy deposition in the scintillator, subtracting the deposition due to ionization which does not contribute significantly to the ${}^9$Li and ${}^8$He yield.   In our next paper we will describe the effect of the ${}^9$Li and ${}^8$He yields on the physics goals of these experiments, which will allow an optimal choice of veto strategies for these experiments given certain assumptions regarding the yet unknown tracking abilities of the detectors.

The unprecedented size of these detectors leads to a second consequence.  As shown in Ref.~\cite{bundle}, interactions of cosmic rays with the atmosphere often yield muon bundles consisting of multiple muons which travel in nearly the same direction.  At the relevant depths the separations of muons are typically of order 10 meters.  This means that in the case of a relatively small detector like KamLAND typically only a single muon will be observed in each event.  As a result less than 5\% of the muon events at KamLAND resulted in the detection of multiple muons \cite{kamlandback}.  On the other hand  JUNO, RENO 50 and LENA are all much larger than 10 meters and so a majority of muon bundles will result in multiple muons impacting the detector.  This will present several challenges.  First, it will be nearly impossible to determine how much energy was deposited by which muon, and so to determine whether one or both muons is showering.  Next, it will be more difficult to track the individual muons, although this tracking is important because full detector cuts in such a large detector weigh heavily upon the detector efficiency.  Third, while we will find that the probability of an individual muon showering is about 20\%, the probability that at least one muon is showering in a two muon event is about 30\%.  In Sec.~\ref{bundlesez} we will estimate the muon bundle frequency in various cases of interest.  

Finally in Sec.~\ref{issez} we will determine the expected rates of the spallation isotopes ${}^9$Li and ${}^8$He.  We will see that these background rates are greater than the expected reactor neutrino signals, and so some veto scheme is necessary.

\section{Showering Muons} \label{docciasez}

We have used the initial muon energy distribution at several depths of interest as the input of a FLUKA simulation which determined the energy deposited by each muon in a spherical (inner) detector containing 20 kton of the liquid scintillator.    As the composition of the outer detector has not yet been fixed, we made the crude approximation that outside of the detector lies a vacuum and so did not consider showers beginning in this region.

Muons lose energy via various processes.  Of these, ionization occurs at an average rate per track length which is independent of muon's energy so long as this energy far exceeds the muon's mass, as it does for muons that traverse the entire detector.  On the other hand, other processes, such as bremsstrahlung, pair production and photonuclear interactions lead to energy dissipation rates which increase with increasing muon energy.  The charged particles produced or liberated in these interactions themselves lose energy via ionization, called secondary ionization.  When we write {\it{ionization}} we will always mean primary ionization, that created by the muon directly.

\begin{figure} 
\begin{center}
\includegraphics[width=4.2in]{mioinit.pdf}
\caption{The energy distribution of single cosmogenic muon events at a depth of 700 meters and 900 meters of rock assuming a flat surface and also at the proposed locations of the experiments JUNO and RENO 50, as well as 200 meters under the location proposed for JUNO}
\label{initfig}
\end{center}
\end{figure}

Most electrons liberated by ionization have a low energy and do not travel far from the track.  However, the energy distribution has a long high energy tail consisting of electrons, called $\delta$ rays, which can travel far from the track.  The former, due to their low energy, do not produce isotopes and so are irrelevant to our backgrounds.  Furthermore they are liberated at such a high rate that the amount of energy lost by the muon by this processes can be calculated quite precisely by simply multiplying the track length by a constant.  In the case of the scintillator relevant for JUNO and RENO 50, this constant is 1.43 MeV/cm.  Therefore, when we refer to the energy deposited minus ionization, we mean the energy deposited minus the track length multiplied by 1.43 MeV/cm.  This is a quantity available to the experimenter assuming that the muon is well tracked.  FLUKA separately calculates the $\delta$ ray production, where $\delta$ rays are defined as primary ionization electrons with energies greater than 100 keV.  Although technically these too are created by ionization, we will include them in our ionization subtracted plots.  In summary, when we state that we subtract the ionization energy we always mean that we subtract 1.43 MeV/cm which to a very good approximation is equal to the expected primary ionization energy not counting the high energy $\delta$ ray tail.  We will refer to the ionization subtracted energy as the showering energy.  

\begin{figure} 
\begin{center}
\includegraphics[width=4.2in]{delta_energy_loss.pdf}
\caption{The energy deposition of $\delta$ rays in the JUNO detector as a function of track length.  The average deposition is roughly 0.88 MeV/cm, while the most common deposition is only about 0.6 MeV/cm.}
\label{deltafig}
\end{center}
\end{figure}

In fact $\delta$ rays, except for perhaps the very far end of their high energy tail, also are not sufficiently energetic to lead to produce isotopes.  Furthermore in a large detector like JUNO, a single $\mu$ creates so many $\delta$ rays that the energy lost due to $\delta$ ray emission can be roughly approximated by a constant energy deposition per track length, although the scatter is nonnegligible as can be seen in Fig.~\ref{deltafig}.  Therefore one may expect the production of spallation isotopes to be correlated with the energy deposition with both the ionization and $\delta$ ray energies subtracted.  In Figs.~\ref{lispectra} and \ref{cumfig} we will therefore consider the total energy, the showering energy which is equal to the showering energy minus the ionization energy and also the showering minus minus the expected $\delta$ ray energy.  The average $\delta$ ray energy deposition is $0.88$ MeV/cm however this is dominated by a long high-energy tail which contains a relatively small portion of the events.  The $\delta$ ray energy deposition from the vast majority of muons is about $0.6$ MeV/cm, as is visible in Fig.~\ref{deltafig}.  Therefore in these figures we subtract not the average $\delta$ ray energy, which is highly sensitive to the long tail, but rather the peak deposition energy $0.6$ MeV/cm. 


The diameter of the detector considered is about 35 meters, which is much larger than typical separation of muons in a muon bundle \cite{bundle}.  As a result if one muon from a bundle strikes the detector, we will see that it is likely that all muons in the bundle strike the detector and so muon bundles lead to multimuon events. 
This is quite different from the case of KamLAND, whose detector diameter is of order a bundle size and so in general at most one muon from each bundle strikes the detector.  As a result the single muon rate at KamLAND includes most bundle events.  Multimuon events will be the subject of Sec.~\ref{bundlesez}.  For simplicity, in this section we will consider only single muon events.

Following \cite{bundle}, the flux of incident muons arriving from a zenith angle $\theta$ is 
\beq
K(h,\theta)=7.2\times 10^{-3}h^{-1.927}\cos(\theta)e^{(-0.581h+0.034)/\cos(\theta)}  \label{keq}
\eeq
where $h$ is the weighted difference in altitude measured in kilometers between the detector and the surface.  The material will always be standard rock, and so we define a weighted altitude $h$ as 2.64 times the true altitude $h_r$.  While this note was written for a flat surface in which case $h$ is a constant, we will also consider detectors under mountains in which case $h$ may be a function of $\theta$.  This is the correct prescription for generalizing the results of \cite{bundle} to nontrivial topographies because muons arising from different angles $\theta$ are independent of one another, and so $h$ can be chosen independently for each value of $\theta$.  

As a result of the factor of $\cos(\theta)$, Eq.~(\ref{keq}) yields the flux of muons crossing a horizontal surface of unit area.  Our detector is not a horizontal surface, it is a sphere.  Therefore the correct angle in our case is not $\theta$ but rather the angle between the normal to the detector and the direction from which the muon arrives.  However we may simplify our expression for the flux by noting that, for a spherical detector, whatever the angle $\theta$ of the incoming muon the total cross sectional area of the detector is $\pi r^2$ where $r$ is the detector's radius.  We set $r$ to 18 meters although the inner detector radius is actually 17.7 meters.  As a result of spill in of the muon showers the fiducial volume for muon events is likely to be greater than 18 meters.  Thus the incident flux upon the detector from an angle $\theta$ is $\pi r^2 K(h,\theta)/\cos(\theta)$ where the factor of $\cos(\theta)$ has been removed as $\pi r^2$ is the area of the detector perpendicular to the velocity of the incident muon.  Integrating this quantity over the angles from which the muons arrive yields the single muon event rate
\beq
R=\pi r^2 \int d\Omega K(h,\theta)/\cos(\theta) = 2 \pi^2 r^2 \int_0^{\pi/2} d\theta K(h,\theta)\tan(\theta) \label{rate}
\eeq  
where we recall that, if the terrain is not flat, $h$ will be a nontrivial function of $\theta$.

The muon flux per unit energy at a given angle $\theta$ is
\bea
\frac{dN(E,h,\theta)}{dE}&=&[E+\epsilon(1-e^{-0.42 h\sec({\theta})})]^{0.232\ln(h)-3.961}\nonumber\\
\epsilon&=&0.0304e^{0.359h}\sec({\theta})-0.0077h+0.659 
\eea
up to a normalization term which can be fixed by demanding that the integral of $N$ over $E$ yield the incident flux $\pi r^2  K(h,\theta)/\cos(\theta)$.   $E$ is the energy in units of TeV.  Fixing this normalization and integrating over solid angles in the upper half of the detector's surface one finds that the muon rate per unit energy is
\bea
\frac{dR(E)}{dE}&=& 2 \pi^2 r^2 \int_0^{\pi/2} d\theta G(h,\theta) K(h,\theta) \frac{dN(E,h,\theta)}{dE}\tan(\theta) \label{enrate}\nonumber\\
G(h,\theta)&=&(2.961-0.232\ln(h))[\epsilon(1-e^{-0.42 h\sec({\theta})})]^{(2.961-0.232\ln(h))}
\eea  
where again we recall that for a nonflat topography $h$ is a function of $\theta$.   The rate per energy is reported Fig.~\ref{initfig} for detectors 700 meters and 900 meters underneath a flat standard rock surface, at the preferred JUNO and RENO 50 sites and also 200 meters below the preferred JUNO site.

\begin{figure} 
\begin{center}
\includegraphics[width=3.5in]{survival_prob.pdf}
\caption{The fraction of muons that traverse the detector as a function of muon energy.  When the energy is less than 10 GeV the constant energy loss per path length caused by primary ionization including $\delta$ ray emission can stop muons inside of the detector, while higher energy muons essentially never stop in the detector.}
\label{sopraviv}
\end{center}
\end{figure}



Given the initial energy and the trajectory of a muon with respect to the detector, FLUKA yields the energy deposited via various channels.  We have simulated muons incident upon the detector with various impact parameters, weighted according to their likelihood such that all muons pass through the inner detector.   As can be seen in Fig. \ref{sopraviv}, unless the muon energy is less than 10 GeV essentially all muons that enter the detector also exit.

\begin{figure} 
\begin{center}
\includegraphics[width=3.5in]{2d_showering.pdf}
\caption{The initial and deposited (not counting ionization) energy of cosmogenic muons at the preferred site for the JUNO experiment.}
\label{initfinal}
\end{center}
\end{figure}

Each datapoint in Fig.~\ref{initfinal} corresponds to the initial energy and the energy deposited not counting ionization of one of our simulated muons.  One may observe that when the muon energy is greater than 10 GeV, the most likely energy deposition is 2 GeV.   In Fig. \ref{initfinaldelta} the this deposition is decomposed into various processes and it can be seen that the 2 GeV maximum is due to the production of $\delta$ rays.  This is consistent with the observation that energy of this maximum is independent of the initial muon energy.   Had bremsstrahlung, pair production or photonuclear interactions instead dominated the deposition they would have led to a peak deposition energy which would have increased with the muon energy. Integrating this distribution over the initial energy one arrives at the upper panel of Fig.~\ref{finalfig}, which displays the distribution $\rho(E)$ of energies deposited by muons at the preferred site for the JUNO experiment and 200 meters deeper.   

Integrating $\rho(E)$ from $E$ to $\infty$ and normalizing the result to unity we obtain the lower panel of Fig.~\ref{finalfig}.  This displays the fraction of muons for which the deposited energy, not counting that deposited by ionization, is greater than each fixed level $E$.  The definition of a showering muon used by the KamLAND collaboration is one which deposits 3 GeV of energy in addition to that deposited by ionization.   As can be seen in Fig 7.18 of Ref.~\cite{dwyer}, this definition is useful as, in the case of a KamLAND sized detector, the vast majority of ${}^9$Li and ${}^8$He is produced by showering muons.  The corresponding results for JUNO are shown in Fig. \ref{lispectra}.  One readily sees that at JUNO and RENO 50 the contribution of nonshowering muons to ${}^9$Li and ${}^8$He production will have little effect on the science goals of the experiment even if they are not vetoed at all.   The lower panel of Fig.~\ref{finalfig} shows that the fractional energy deposition distribution of the muons is quite insensitive to the depth in the range considered and in fact the showering fraction is extremely robust.  For a 20 kton detector using KamLAND's definition of a showering muon, as had been anticipated in Ref.~\cite{menufact} about {\bf{20\% of muons are showering at these depths}}, 200 meter variations in the depth and indeed even 200 meter changes in the topography have little effect on this robustly determined fraction.    This is our main result.

\begin{figure} 
\begin{center}
\includegraphics[width=3in]{inelastic.pdf}
\includegraphics[width=3in]{deltaRay.pdf}
\includegraphics[width=3in]{brem.pdf}
\includegraphics[width=3in]{pairProd.pdf}
\caption{As in Fig. \ref{initfinal} but now decomposed into various processes.  The main inelastic process consists of photonuclear interactions.  One can observe that $\delta$ ray emission is responsible for most energy deposition and in particular for the spectrum of the energy deposition.  These deposit a constant energy per path length of the track.  As seen in Fig. \ref{sopraviv}, when a muon's energy is less than about 10 GeV  it cannot cross the detector and so the energy deposited is proportional to the track length which is proportional to the muon energy.}
\label{initfinaldelta}
\end{center}
\end{figure}

\begin{figure} 
\begin{center}
\includegraphics[width=in]{1d_showering.pdf}
\includegraphics[width=4in]{efficiency.pdf}
\caption{Top panel: The rate of muons depositing a given amount of energy is shown, not including the energy deposited by ionization, at the preferred site for the JUNO experiment and 200 meters deeper.  Bottom panel: The normalized integral of the top panel, showing the fraction of muons which deposit more than a given threshold energy.  Note that only the high energy tail of this curve is sensitive to the overhead burden in this range.  The green line illustrates that 20 percent of muons deposit more than 3 GeV in addition to that deposited by ionization.}
\label{finalfig}
\end{center}
\end{figure}

\section{Muon Bundles} \label{bundlesez}

In this section, we will use the parametric formulae of Ref.~\cite{bundle} to determine the multimuon event rates at 20 kton spherical liquid scintillator detectors in various settings. We will then calculate the probability that more than a muon of a single bundle will hit the detector.  The $m$-muon flux is
\beq
\Phi(h,\theta,m)=\frac{K(h,\theta)}{m^{(-0.0771h^2+0.524h+2.068)e^{0.03e^{0.47h} \sec(\theta)}}}. \label{phieq}
\eeq
The $m$-muon rate can be calculated as in Eq.~(\ref{rate})
\beq
R(m) = 2 \pi^2 r^2 \int_0^{\pi/2} d\theta \Phi(h,\theta,m)\tan(\theta).
\eeq  
This is the muon bundle rate, which is the rate with which the axes of muon bundles strike the detector. To relate this number to an observable quantity, one can observe that $mR_m$ is the rate at which muons which are part of $m$-muon bundles strike the detector.  Here $m$ is the total number of muons in the bundle, whether or not they all strike the detector.

Of course, one is interested not in the number of muons in a bundle but in the number of muons which actually strike the detector.  We will now calculate this in the case of 2-muon bundles, while in the case of bundles with 3 or more muons we will make the crude approximation that at least 2 always strike the detector.  Ref.~\cite{bundle} provides the probability density $f(R)=dN/dR$ of the separation $R$ between a muon in a bundle and the axis of the shower that generated the bundle 
\beq
f(R)=C\frac{R}{(R+R_0)^\alpha}
\eeq
where 
\bea
\alpha(h,M)&=& (-0.448M+4.969)e^{(0.0194M+0.276)h} \nonumber \\ 
 R_0(h,\theta,M)&=&\frac{\alpha(h,M)-3}{2}\frac{(-1.786 M +28.26)h^{-1.06M}}{e^{10.4(\theta-1.3)}+1}
 \nonumber\\
 C&=&(\alpha-1)(\alpha-2)R_0^{\alpha-2}.
\eea
Below we will need the probability density not for the distance from the axis, but rather for the distance $D$ between two muons
\beq
f'(D)=\frac{dN}{dD} .
\eeq
Consider a 2 muon bundle and let $R$ and $R'$ be the distance between the two muons and the bundle axis. $R'$ can be obtained from the distance $D$ and the angle $\phi$ between the two muons
\beq
R'(R,D,\phi)=\sqrt{R^2+D^2-2RD\textrm{Cos}(\phi)} .
\eeq
In order to calculate $f'(D)$, we consider the distribution of the muons per element of area normal to the axis
\beq
\frac{dN}{dA}=\frac{f(R)}{2\pi R}.
\eeq
The distribution $f'(D)$ is then the integral
\beq
f'(D)=\int_{R=0}^{\infty}\int_{\phi=0}^{2\pi} f(R) f(R'(R,D,\phi))\frac{D}{2\pi R'(R,D,\phi)} \textrm{d}R \textrm{d}\phi .
\eeq
The probability that the detector is hit by both muons in a 2-muon bundle is
\beq
P_{M\mu}=\int_{b=0}^{R_d}\int_{\theta=0}^{\pi/2} \left(\int_{D=0}^{R_d-b}f'(D) \textrm{d}D + \int_{D=R_d-b}^{R_d+b}f'(D)\textrm{arccos}\left(\frac{b^2 + d^2 - R_d^2}{2bD}\right) \textrm{d}D \right)\textrm{d}b \textrm{d}\theta
\eeq
where $R_d$ is the radius of the detector.  We evaluate this quantity numerically for each candidate site.

In Table~\ref{ratetab} we summarize $P_{M\mu}$ together with several other relevant quantities.  The muon rate is defined to be $\sum_{m=1}^{\infty} mR(m)$.  The quantity $\sum_{m=1}^{\infty} R(m)$ is the sum of the single muon and bundle rates.  This systematically underestimates the event rate by about 3-5\% because, as we have seen, sometimes only some of the muons from a given bundle strike the detector.   However this correction is neglible compared to our uncertainties, which are around 20\%, and so we will simply refer to this quantity as the event rate.  The bundle rate is $\sum_{m=2}^{\infty} R(m)$. The $m$-muon rate is just $R(m)$ while the mean muon energy is calculated for single muons with energy below about 5 TeV.  The latter is higher than typical values given in the literature as a result of the long high energy tail which consists of few muons but leads to a significant fraction of the isotope production.  

We have considered a detector 700 and 900 meters underneath a flat surface as well as various topographies corresponding to potential sites for detectors.  In particular we have considered the preferred Dong Keng \cite{noisim} site for JUNO and the preferred Mt GuemSeong \cite{noisim} site for RENO 50, which are illustrated in the upper and lower panels of Fig.~\ref{mappe} respectively.  These are respectively $h_r=700$ meters and $900$ meters underneath the peaks of their corresponding hills, where $h_r$ is the rock overburden which is equal to $h_r=h/2.64$.  We have also considered a location 200 meters beneath the preferred site for JUNO.   Note that the reported mean muon energies are higher than that found in other studies.  This increase is caused by a small number of muons in the very high energy tail of the cosmogenic muon distribution, at several TeV.  

To determine the expected muon flux ideally one requires a geological survey of the rocks around these sites and a full 3d simulation such as MUSIC.   We have instead simply assumed that the rock is standard and have employed a cylindrically-symmetric approximation to the topographies illustrated in Fig.~\ref{mappe}.  For JUNO we have assumed an overburden of $h_r=700$ meters of rock in a 100 meter radius circle, followed by an annulus of inner radius 100 meters and outer radius 500 meters with a surface 50 meters lower, than another annulus of outer radius 700 meters with a surface 50 meters yet lower and finally we have assumed that the surface is another 50 meters lower at radii beyond 700 meters.  Similarly we have approximated Mt Guemseong using a series of annuli, beginning with a 200 meter radius circle with $h_r=900$ meters of rock overburden, followed by a 600 meter annulus with $h_r=800$ meters of overburden, a 900 meter radius annulus with $h_r=700$ meters of overburden and then we have assumed a $h_r=600$ meter overburden at all radii beyond 900 meters.  

\begin{figure} 
\begin{center}
\includegraphics[width=4.2in]{JUNO-hires.jpg}
\includegraphics[width=4.2in]{reno-hires.jpg}
\caption{Google maps illustrating the favored locations of the JUNO and RENO 50 detectors}
\label{mappe}
\end{center}
\end{figure}

\begin{table}[position specifier]
\centering
\begin{tabular}{c|l|l|l|l|l}
&700 m&900 m&JUNO&JUNO+200m&RENO 50\\
\hline\hline
muon rate&$3.0\pm 0.7$&$1.1\pm 0.2$&$5.4\pm 1.2$&$1.9\pm 0.4$&$3.1\pm 0.6$\\
\hline
event rate&$2.3\pm 0.5$&$0.90\pm 0.19$&$4.1\pm 0.9$&$1.5\pm 0.3$&$2.4\pm 0.5$\\
\hline
bundle rate&$0.36\pm 0.09$&$0.12\pm 0.03$&$0.69\pm 0.16$&$0.22\pm 0.05$&$0.37\pm 0.08$\\
\hline
single $\mu$ rate&$2.0\pm 0.4$&$0.78\pm 0.16$&$3.4\pm 0.7$&$1.3\pm 0.3$&$2.1\pm 0.4$\\
\hline
two $\mu$ rate&$0.23\pm 0.06$&$0.081\pm 0.019$&$0.43\pm 0.10$&$0.14\pm 0.03$&$0.24\pm 0.05$\\
\hline
$P_{M\mu}$& $0.57$ & $0.65$ & $0.51$ & $0.60$ &$0.52$\\
\hline
three $\mu$ rate&$0.067\pm 0.017$&$0.022\pm 0.005$&$0.13\pm 0.03$&$0.040\pm 0.010$&$0.069\pm 0.016$\\
\hline
mean $\mu$ energy&$267\pm 8$&$310\pm 8$&$254\pm 7$&$294\pm 7$&$284\pm 7$\\
\end{tabular}
\caption{Rates in Hz of various kinds of events under 700/900 meters of rock with a flat surface, at JUNO, 200 meters beneath JUNO and at RENO 50 and the mean energy in GeV.  The errors do not include systematic errors in the neutrino model, but consist of an uncertainty of an overall 30 meters in the surface elevation and a 5\% uncertainty in the rock density.  The $n$ muon rate is the rate of muons striking the detector which are in $n$-muon bundles, regardless of how many of the other $n-1$ muons actually enter the detector.  The quantity $P_{M\mu}$ is the probability that both muons in a 2-muon bundle hit the detector.}
\label{ratetab}
\end{table}

In Table~\ref{ratetab} one can see that at JUNO one expects that for 51\% of all 2-muon events in which a single muon strikes detector, both muons will strike the detector.  As 2-muon events occur at a rate of 0.43 Hz, this means that they contribute 0.22 Hz to the bundle rate.  Events with more than 2 muons occur at a rate of 0.26 Hz.  If we make the crude approximation that in the case of all such events, at least 2 muons enter the detector then we find a multi-muon event rate of 0.48 Hz which is 11\% of the total event rate of 4.3 Hz, where we have added 0.2 Hz corresponding to the fact that 49\% of 2-muon bundles appear as two events, with one muon from each of two bundles.  However although only 11\% of events are multimuon events, these account for 1.7 Hz of the 5.4 Hz total muon rate and so 31\% of all muons.   Thus one expects about 31\% of all cosmogenic muon isotope backgrounds to result from multimuon events, where localized vetoes require very difficult tracking.

\section{Isotope Production} \label{issez}

In Fig. \ref{lispectra} we present the spectra of deposited total energy and deposited showering energy expected at JUNO.  We also plot the deposited showering energy minus the most likely deposited $\delta$ ray energy, which is $0.6$ MeV/cm multiplied by the track length.  In the lower panel we show the ${}^9$Li rate as a function of deposited total energy, showering energy and $\delta$ ray subtracted total energy.    In Fig.~\ref{cumfig} we plot, in the top panel, the fraction of the muons whose energy deposition exceeds a given threshold and in the bottom panel one sees the fraction of ${}^9$Li produced by these muons.  This plot is obtained by integrating and renormalizing Fig.~\ref{lispectra}.

One can again see that a 3 GeV cut on the showering energy would require a veto after about 23\% of all muon events, which if longer than the ${}^9$Li half life would lead to a large dead time.  It would however reject about 95\% of ${}^9$Li production.  On the other hand, a cut near 10 GeV would require a veto of only 3\% of events but only reject 78\% of the ${}^9$Li.  A veto based on the total energy deposition appears mildly more problematic, a 95\% rejection efficiency requires a veto of the 35\% of all muons with total energy deposition above 6.9 GeV.  If the most common $\delta$ ray energy at the corresponding track length is subtracted from the showering energy, one obtains a 95\% rejection efficiency at the price of vetoing the 24\% of all muon events with a $\delta$ ray subtracted showering energy beyond 1.2 GeV.   This leads to a dead time with fixed rejection efficiency which agrees, to within the precision introduced by our binning, with that obtained using a showering energy cut and no $\delta$ ray subtraction.

\begin{figure} 
\begin{center}
\includegraphics[width=4.0in]{combined_noncumulative.pdf}
\caption{Top: The spectrum of the deposited total energy, showering and showering minus expected $\delta$ ray energy at JUNO.  Bottom:  The rate of ${}^9$Li production as a function of the total deposited, showering energy and $\delta$ ray subtracted showering energies.  Note that most ${}^9$Li is produced by the very high energy tail of the energy deposition (above 10 GeV) and so is easily cut.}
\label{lispectra}
\end{center}
\end{figure}

\begin{figure} 
\begin{center}
\includegraphics[width=6in]{combined_cumulative.pdf}
\caption{Top: The fraction of muons for which the deposited energy exceeds the value in the horizontal axis.  Bottom: The fraction of ${}^9$Li produced by cosmogenic muons which deposit more energy than the value on the horizontal axis.  The blue curves are defined by identifying the horizontal axis with the total deposited energy, the red curves use the showering energy while the green curves use the showering energy minus the expected $\delta$ ray deposition given a fixed track length.}
\label{cumfig}
\end{center}
\end{figure}

\begin{figure} 
\begin{center}
\includegraphics[width=4.0in]{lithium_raw.pdf}
\includegraphics[width=4.0in]{helium_raw.pdf}
\caption{The mean number of ${}^9$Li (top) and ${}^8$He (bottom) created by a single $\mu^-$ (red) and $\mu^+$ (blue) with a given energy.  The lines are the best power law fits. The gray band represents the relative statistical uncertainty on the simulated isotope yield.}
\label{liperenergia}
\end{center}
\end{figure}

We have also estimated the total yield of ${}^9$Li and ${}^8$He at each experimental site.  To do this, we ran $10^7$ FLUKA simulations of single monoenergetic muons  of each energy between 1 GeV and 10 GeV and $10^6$ simulations at energy between 11 GeV and 500 with a step size of 1 GeV, with an additional $10^6$ simulations at each energy up to 3 TeV with a step size of 50 GeV, for a total of $6.4 \times 10^8$ simulations.  Again the impact parameter was randomized with a distribution reflecting a homogeneous distribution of muons.  The   ${}^9$Li yield and best power law fit, with an exponent of $0.842\pm 0.002$ are presented in red in the top panel of Fig.~\ref{liperenergia}.  The results of another $6.4\times 10^8$ simulations of antimuons are presented in blue, together with a power law fit whose exponent is $0.847\pm 0.002$.  The same simulations also determined the ${}^8$He rate, as is shown on the bottom panel.  The best fit exponents are $0.869\pm 0.006$ and $0.861\pm 0.006$ for $\mu^-$ and $\mu^+$ respectively.  

The power law fits reproduce the simulated isotope yields to within the statistical errors from 10 GeV up to 3 TeV, where the vast majority of the spallation isotope production occurs.  However below 10 GeV, the isotope production from $\mu^-$ events increases.  Indeed in this range the ${}^9$Li yield is 3.3 times greater for $\mu^-$ events than for $\mu^+$.  This is because $\mu^-$ at these low energies stop and bind with the hydrogen and carbon in the scintillator.  The deeper potential well about the carbon nucleii means that any $\mu^-$ which initially bind to hydrogen are soon transfered to carbon and these quickly decay to 1s or 1p orbitals.  Here most of the $\mu^-$, like the $\mu^+$, simply decay.  However 7\% of the $\mu^-$ are then captured by the carbon nucleii, undergoing a charged current interaction which can create spallation isotopes.

\begin{figure} 
\begin{center}
\includegraphics[width=4.0in]{li_merged_spectra.pdf}
\includegraphics[width=4.0in]{he_merged_spectra.pdf}
\caption{The ${}^9$Li (top) and ${}^8$He (bottom) production rates per energy of the cosmogenic muon at the five experimental sites}
\label{li}
\end{center}
\end{figure}

To arrive at the rate for each experimental site we have folded the power law fits to the simulated data as a function of muon energy with the expected muon spectra reported in Fig.~\ref{initfig}.  In contrast with Fig.~\ref{initfig}, here we are interested in the total $\mu$ rate, not just those arising from single $\mu$ events, so we have rescaled this result by the ratio of the total $\mu$ rate to the single $\mu$ rate reported in Table~\ref{ratetab}.  We have assumed that the ratio of $\mu^+$ to $\mu^-$ is $1.37$, as was found for 1.2 TeV muons by Kamiokande in Ref.~\cite{antimu}.  The resulting ${}^9$Li and ${}^8$He rates per muon energy bin are displayed in Fig.~\ref{li}.  Next, to arrive at the total ${}^9$Li and ${}^8$He rates we have integrated this figure over the $\mu$ energy.  The resulting total rates at each experimental site are summarized in Table~\ref{litab}.  The  ${}^9$Li  rates are compatible with the rough estimate of Ref.~\cite{menufact} obtained by simply rescaling KamLAND's rate.

As can be seen in Table~\ref{litab}, the total spallation isotope rates are appreciably higher than the expected reactor neutrino IBD signal rates of $3\times 10^{-4}$ Hz for JUNO and $2\times 10^{-4}$ Hz for RENO 50 at Guemseong.  However only those decays which produce a neutron yield a false double coincidence signal.  These are $51\%$ of ${}^9$Li decays and only 16\% of ${}^8$He decays.  In the bottom two rows of Table~\ref{litab} we report the false double coincidence rate expected.  Note that this background double coincidence rate is still three times greater than the IBD signal rate at JUNO and RENO 50, although it would be only slightly greater than the RENO 50 signal rate at the Munmyeong site of Ref.~\cite{2rivelsim} even with the same overhead burden as the Guemseong site.

In practice the background rate will be reduced, at the expense of dead time, by a veto program.  Similarly, the expected background can be subtracted from a signal by a shape analysis.  However the statistical fluctuations created by the background will survive this subtraction and obscure the low energy oscillations in the reactor neutrino spectrum whose observation is necessary for a determination of the hierarchy.  Thus, both the sensitivity to the hierarchy and the precision with which $\theta_{12}$ may be measured depend critically on the optimal choice of muon vetoes.  As the muon rate is comparable to the ${}^9$Li  half life, full detector vetoes are not an option for all but the highest energy showers, thus excellent muon tracking, including the tracking of multimuon events and even muons that arrive horizontally will be necessary to achieve these science goals.  

\begin{table}[position specifier]
\centering
\begin{tabular}{l|c|c|c|c|c}
                    &       700 m           &  900 m            &  JUNO             &  JUNO+200m     & RENO 50\\
\hline\hline
${}^9$Li  rate      &       $93\pm20        $&$ 39.3\pm 8.2     $&$ 167\pm 37       $&$ 68\pm 14     $&$ 96\pm 20$\\
\hline
${}^8$He rate       &       $10.3\pm 2.1    $&$ 4.37\pm 0.92    $&$ 18.5\pm 4.1     $&$ 7.5\pm 1.5   $&$ 10.5\pm 2.1$\\
\hline
${}^9$Li  $n$-decay rate&   $47\pm 10       $&$ 20.0\pm 4.2     $&$ 85\pm 19        $&$ 34.6\pm 7.3  $&$ 49\pm 10$\\
\hline
${}^8$He $n$-decay rate&    $1.65\pm 0.35   $&$ 0.70\pm 0.15    $&$ 2.96\pm 0.65    $&$ 1.20\pm 0.24 $&$ 1.69\pm 0.35$\\
\hline
\end{tabular}
\caption{Rates in $10^{-5}$Hz of ${}^9$Li and ${}^8$He production via cosmogenic muon spallation on ${}^{12}$C under 700/900 meters of rock with a flat surface, at JUNO, 200 meters beneath JUNO and at RENO 50.  The resulting ${}^9$Li and ${}^8$He decays only provide false double coincidence signals when their decay produces a neutron, which happens in $51\%$ and $16\%$ of their decays respectively.  The last two rows report the rates of these decays.   The errors reported reflect only the uncertainty in the muon rate, not the uncertainties in the isotope production per muon.}
\label{litab}
\end{table}

\section{Concluding Remarks}

As can be seen in the lower panel of Fig.~\ref{finalfig}, for any depth in the range relevant to future large liquid scintillator experiments, about 20\% of single muons will be showering in a 20 kton spherical detector.  Thus to determine the showering single muon rate, one only needs to find the single muon rate for a given site and multiply it by 20\%.  Similarly from Table~\ref{ratetab} we can see that 17-20\% of muon events will occur in muon bundles.  As, for the large detectors considered here, in most cases muon bundles yield multimuon events in the detector, the multimuon event rate is about 11\%.   Again, this bundle fraction is fairly robust against changes in depth and topography and so may be applied to a wide variety of candidate experimental sites.

One immediate consequence of our study is that both the bundle rates are roughly 0.8 Hz (0.5 Hz) at JUNO (RENO 50) with the shower rates only about 40\% lower.  This means that KamLAND's veto strategy, employing 2 second full detector cuts for showering and poorly constructed tracks, cannot be applied to these experiments as the events are separated by less than 2 seconds.  If JUNO is moved 200 meters lower, KamLAND's cuts would still be problematic although 1 second full detector cuts would be possible with considerable dead time.  1 second cuts are sufficient to reduce the ${}^9$Li and ${}^8$He backgrounds well below the level of the signal.  KamLAND's veto strategy has been assumed in all studies of these backgrounds that have so far appeared in the literature, and so our results indicate that these studies need to be redone.

As a result, full detector cuts will be infeasible at these experiments and it is critical that muon tracking be successful for the vast majority of single and multimuon events, so that selective cuts may remove events from cylinders surrounding cosmogenic muon tracks.  A muon veto system above the main detector will help with downgoing muons.  However, a flat 1300 m${}^2$ detector placed 5 meters above the JUNO (RENO 50) inner detector will only be exposed to 40\% (37\%) of the muons that eventually pass through the inner detector and these will in general be somewhat less energetic and so produce less isotopes than average.  

In our next paper we will investigate the creation of ${}^9$Li and ${}^8$He and determine the effectiveness of various veto schemes  as well as the cost to the scientific goals both as a result of the increased dead time and as a result of the background which will mask the 1-3 oscillations in the 3-4 MeV range whose detection is essential to determine the neutrino mass hierarchy \cite{noiteor}.  However for any proposed veto scheme, the lower panel of Fig.~\ref{finalfig} together with the total muon flux may be used to calculate the resulting veto efficiency and so the resulting dead time for the experiment.  

\section* {Acknowledgement}
\noindent
We are pleased to thank Xin Qian and Anton Empl for suggestions and correspondence.  JE is supported by NSFC grant 11375201 and a KEK fellowship.  EC  is supported by the Chinese Academy of Sciences Fellowship for Young International Scientists grant  number 2013Y1JB0001 and NSFC grant 11350110500.  XZ is supported in part by  NSFC grants 11121092, 11033005 and 11375202.   


\bibitem{fiducia}
  E.~Ciuffoli, J.~Evslin and X.~Zhang,
  ``Confidence in a neutrino mass hierarchy determination,''
  JHEP {\bf 1401} (2014) 095
  [arXiv:1305.5150 [hep-ph]].
 J.~Evslin,
  ``Confidence in the neutrino mass hierarchy,''
  arXiv:1310.4007 [hep-ph].

\end{document}

\bibitem{2rivel}
   E.~Ciuffoli, J.~Evslin and X.~Zhang,
  JHEP {\bf 1212} (2012) 004
  [arXiv:1209.2227 [hep-ph]].
  E.~Ciuffoli, J.~Evslin, Z.~Wang, C.~Yang, X.~Zhang and W.~Zhong,
  arXiv:1211.6818 [hep-ph] and
  arXiv:1308.0591 [hep-ph].

\bibitem{daed}
  J.~Alonso, F.~T.~Avignone, W.~A.~Barletta, R.~Barlow, H.~T.~Baumgartner, A.~Bernstein, E.~Blucher and L.~Bugel {\it et al.},
  arXiv:1006.0260 [physics.ins-det].

\bibitem{lena}
 M.~Wurm {\it et al.}  [LENA Collaboration],
  Astropart.\ Phys.\  {\bf 35} (2012) 685
  [arXiv:1104.5620 [astro-ph.IM]].

\bibitem{whitepaper}
  C.~Aberle, A.~Adelmann, J.~Alonso, W.~A.~Barletta, R.~Barlow, L.~Bartoszek, A.~Bungau and A.~Calanna {\it et al.},
  arXiv:1307.2949 [physics.acc-ph].

\bibitem{lsnd}
  A.~Aguilar-Arevalo {\it et al.}  [LSND Collaboration],
  Phys.\ Rev.\ D {\bf 64} (2001) 112007
  [hep-ex/0104049].

\end{thebibliography}

\end{document}